\documentclass[a4paper,11pt]{article}
\usepackage{graphicx}
\usepackage{epsf,amsmath,bbold,amsfonts,stmaryrd}

\usepackage[utf8]{inputenc}
\usepackage{mathrsfs}
\usepackage{appendix}
\usepackage{amssymb}
\usepackage{float}
\usepackage{color}
\usepackage{cite}
\usepackage{hyperref}
\hypersetup{pageanchor=false}
\usepackage{indentfirst}
\usepackage{url}
\usepackage{float}
\usepackage{caption}
\usepackage[numbers,square,comma,sort&compress,merge]{natbib}

\def\a{\alpha}
\def\b{\beta}

\def\D{\Delta} 
\def\f{\frac}
\def\g{\gamma}
\def\gmn{g _ {\m \n}}

\def\lm{\lambda} %
\def\l{\left}

\def\m{\mu} %
\def\n{\nu} %
\def\nn{\nonumber}

\def\p{\phi} %
\def\td{\tilde} %
\def\r{\right}

\def\S{\Sigma}
\def\t{\theta}
\def\T{\Theta}

\def\z{\zeta}

\def\be{\begin{equation}}
\def\ee{\end{equation}}
\def\bag{\begin{aligned}}
\def\eag{\end{aligned}}
\def\bea{\begin{eqnarray}}
\def\eea{\end{eqnarray}}
\def\ba{\begin{array}}
\def\ea{\end{array}}

\def\bc{\begin{center}}
\def\ec{\end{center}}
\def\bl{\begin{flushleft}}
\def\el{\end{flushleft}}
\def\br{\begin{flushright}}
\def\er{\end{flushright}}
\def\bi{\begin{itemize}}
\def\ei{\end{itemize}}

\def\bt{\begin{tabular}}
\def\et{\end{tabular}}

\numberwithin{equation}{section}

\newcommand{\pc}[1]{\textcolor{black}{}}

\numberwithin{equation}{section}

\begin{document} 

\title{\textbf{Revisiting the shadow of braneworld black holes}}
\author{
Yehui Hou$^{1}$, Minyong Guo$^{1*}$,
Bin Chen$^{1,2,3}$}
\date{}

\maketitle

\vspace{-10mm}

\begin{center}
{\it
$^1$Center for High Energy Physics, Peking University,
No.5 Yiheyuan Rd, Beijing 100871, P. R. China\\\vspace{1mm}

$^2$Department of Physics, Peking University, No.5 Yiheyuan Rd, Beijing
100871, P.R. China\\\vspace{1mm}

$^3$ Collaborative Innovation Center of Quantum Matter,
No.5 Yiheyuan Rd, Beijing 100871, P. R. China\\\vspace{1mm}
}
\end{center}

\vspace{8mm}

\begin{abstract}
We revisit the shadows of rotating braneworld black holes in the Randall-Sundrum type II model, by considering not only the metric in the near region of the black hole but also the linearized metric in  the far region where the observer stays. Our study is significantly different from previous studies, which relies only on the metric in the near region. From the study, we identify a critical angle $\theta_c$ which decides the shadow curve is open or closed:  the shadow curve would be open if the observational angle $\theta_c<\theta_o\le\pi/2$, and the curve becomes closed when $0<\theta_o\le\theta_c$. We study how various parameters affect the shape of the shadow curve. We furthermore apply our analysis to the data of M87* from the Event Horizon Telescope and obtain a new constraint on the parameters of the braneworld black holes. 
\end{abstract}

\vfill{\footnotesize Email:yehuihou@pku.edu.cn,\,minyongguo@pku.edu.cn,\,bchen01@pku.edu.cn\\$~~~~~~*$ Corresponding author.}

\maketitle

\newpage
\section{Introduction} 
In recent years, the discovery of gravitational waves from LIGO and Virgo \cite{Abbott:2016blz, TheLIGOScientific:2017qsa, Abbott:2016nmj} and the image of the supermassive black hole at the center of M87 taken by the Event Horizon Telescope (EHT) \cite{Akiyama:2019cqa, Akiyama:2019sww, Akiyama:2019brx, Akiyama:2019bqs, Akiyama:2019fyp, Akiyama:2019eap} have brought the study of black holes into a new era, in which the theoretical studies can be in accord with accurate measurements. It is now feasible to probe the most dynamic regions in  the universe, to investigate various long-standing problems in astrophysics, and to test  gravitational theories in the strong field area.

As we know, Einstein's general relativity (GR) has been considered by far the most successful theory of gravity. And  in many studies it is usually assumed that astrophysical black holes are described by the Kerr metric which is the most important solution in GR. However, as argued by \cite{BSS}, even though gravitational waves and black holes shadows can be well explained by GR, there are still good reasons to explore alternative theories of gravity beyond GR. These reasons include but are not limited to the problem of the singularities in black holes \cite{Christodoulou:1991yfa} and in cosmology \cite{Penrose:1964wq, Hawking:1969sw}, the origin of dark matter \cite{Milgrom:2003ui}, the nature of dark energy \cite{Perlmutter:1998np, Peebles:2002gy}, and so on. On the other hand, most modified gravity theories   often contain extra degrees of freedom, which have significant physical implications \cite{Nojiri:2017ncd, Maartens:2003tw, Kanti:2004nr}. Consequently they get strong constraints from the experiments, say the classical tests in the solar system,  and the observations on cosmological expansion, CMB etc.. The recent astronomical observations including gravitational waves  and black hole shadows can impose new constraints on the parameters in these theories as well\cite{Akiyama:2019cqa, Akiyama:2019sww, Akiyama:2019brx, Akiyama:2019bqs, Akiyama:2019fyp, Chakravarti:2019aup}. In particular, some interesting results related to the black hole shadows have been found. For examples, one can see the black hole shadows in the regularized $4D$ Einstein-Gauss-Bonnet gravities in \cite{Guo:2020zmf,Konoplya:2020bxa,Kumar:2020owy,Wei:2020ght}, the shadows of black holes in an expanding universe in \cite{Bisnovatyi-Kogan:2019wdd,Bisnovatyi-Kogan:2018vxl,Li:2020drn}, the cuspy and fractured black hole shadows in\cite{Cunha:2017eoe,Qian:2021qow}, the black holes furnished by non-geodesic photons in \cite{Chen:2020qyp,Hu:2020usx} and the novel shadows from the asymmetric thin-shell wormhole in \cite{Wang:2020emr,Wielgus:2020uqz, Tsukamoto:2021fpp, Guerrero:2021pxt, Peng:2021osd}.

Thereinto,  the braneworld  scenario is of particular interest, not only in cosmology but also in particle physics and string theory. It assumes that we live in  four dimensional timelike hypersurface of a fundamentally higher dimensional spacetime. The existence of extra dimensions have many interesting implications in particle physics and cosmology. It would be certainly interesting to study its implication on black hole shadow. In \cite{Amarilla:2011fx}, the authors studied the shadow cast by a rotating  black hole in the Randall-Sundrum(RS) braneworld. In \cite{Eiroa:2017uuq}, they generalized the study to involving a cosmological constant. In \cite{Abdujabbarov:2017pfw}, the gravitational lensing and retrolensing in a braneworld black hole spacetime has been investigated. Recently, in \cite{BSS}, the authors used the data of the image of M87* to confine the parameters of the rotating braneworld black hole. After taking a closer look at these studies, we find that all of them relied exclusively  on the metric which is only valid near the black hole and cannot be extended to the far region. This is unsatisfying as the image is taken with respect to the observer in the far region. When one  moves far away from the black hole, the spacetime should be described by the linearized, weak field form of the metric \cite{RWW}. In other words, in order to obtain the shadow curves in the RS model, one must take into account the effects of the linearized metric. Based on this motivation, we investigate the shadow curves of the rotating braneworld black holes by considering both the metric in the near region and the linearized metric in the far region. Then, we apply our analysis to the black hole shadow of M87 and obtain a new constraint on the parameters of braneworld black holes, which differs significantly from the one found in \cite{BSS}.

This paper is organized as follows. In sec. \ref{section2} we study the null geodesics in the near region and the far region, respectively. The two null geodesics can be related to each other by the conserved charges, which can be defined with respect to the same Killing symmetries in two different regions. In sec. \ref{shadow}, we study the shadow curves in the sky of distant observers. Then in sec. \ref{sec4} we apply the previous results to the image of M87*. We close this paper with a summary in sec, \ref{sec5}. 

\section{Null geodesics in RS braneworlds}\label{section2} 

In this section, we give a brief review on the rotating black holes in RS type-II braneworld. As far as we know, there is no complete black hole solution covering the whole braneworld. However, an  approximate solution has been found to portray the near region of the braneworld black hole, which is the so-called tidal Kerr black hole solution. In contrast, in the  region far away from the black hole, the spacetime is known to be governed by a linearized weak field metric \cite{RWW}. Therefore, we begin to investigate the null geodesics in the near and the far regions, respectively. Then we connect these geodesics with the conserved quantities determined by the symmetries of the spacetime which must remain the same in the whole braneworld. 

\subsection{Null geodesics in the near region} 

Let us start with the null geodesics in the near region where the form of the braneworld solution is the same to the Kerr-Newman  solutions in the Einstein-Maxwell theory \cite{Aliev:2005bi, Aliev:2009cg}, that is,
\bea\label{ns}
ds^2 =&-&\l(1-\f{2Mr-b}{\S}\r)dt^2 -2\f{a\l(2Mr-b\r)\sin^2{\t}}{\S}dtd\phi + \f{\S}{\D}dr^2 + \S d\t^2\nn \\
&+& \l( r^2 + a^2 +\f{2Mr-b}{\S}a^2\sin^2{\t} \r)\sin^2{\t}d\p^2,
\eea
where
\bea
\D &=& r^2 -2Mr +a^2 +b \ ,\nn \\
\S &=&r^2 + a^2\cos^2{\t} \ .
\eea
$M$ is the mass, $a = J/M$ is the angular momentum parameter which can be set to $a \geq 0$ without loss of generality. The parameter $b$ is called the tidal charge,  representing the imprint of non-local gravitational effects from the bulk spacetime. In contrast to the square of electric charge $q^2$ in the Kerr-Newman metric, the charge $b$ can be both positive and negative. A negative $b$ may strengthen the gravitational field \cite{Dadhich:2000am}. The event horizon is located at $\D = 0$, which gives
\be
r_h^{\pm} = M \pm \sqrt{M^2 - (a^2 + b)} \ .
\ee
In the following, we use $r_h$ to represent the outer event horizon for short. If the weak cosmic censorship conjecture holds, there is no naked singularity,  then the condition $a^2 + b \le M^2$ is required. Thus, the black hole spin parameter $a$ is not limited to be smaller than the mass $M$, since $b$ can be negative. In addition, It is worth mentioning that the metric (\ref{ns}) cannot be extended to  the far field region,  and it cannot describe the entire spacetime of a rotating black hole on the brane.

Next, let us consider massless particles moving along null geodesics in the near region. The spacetime has a timelike Killing vector field and a rotational Killing vector field. Correspondingly there are two  conserved quantities
\bea
\bag
E &= -p_t \ ,\ L = \lm E = p_{\p} \ , \\
\eag
\eea
in terms of which the Hamilton-Jacobi equation $H =\f{1}{2} \gmn p^{\m}p^{\n}= 0$ can be separated as 
\bea
p_{\t}^2 + \l( a\sin{\t} - \f{\lm}{\sin{\t}} \r)^2 = -\D p_r^2 + \f1{\D}(r^2+a^2-a\lm)^2 = K_n \ .
\eea
Here we have rescaled the affine parameter of null geodesics to set $E=1$, and  the constant $K_n$ is related to the Carter constant $\eta$ by $K_n=\eta+(\lm-a)^2$. Usually, $\eta$ and $\lambda$ are referred to as the impact parameters in the literatures. Then the radial and angular equations of motion read
\bea
\frac{\S}{E} p^r = \pm_r \sqrt{R_n(r)} \ , \quad \frac{\S}{E} p^{\t} = \pm_{\t} \sqrt{\T_n({\t})} \ ,
\eea
~\\
where
\bea
R_n(r)&=&(r^2 + a^2 - a\lm)^2 - \D(r)(\eta + (\lm -a)^2) \ ,\\
\T_n({\t})&=&\eta + a^2\cos^2{\t} - \lm^2\cot^2{\t} \ ,\label{Tn}
\eea
can be regarded as the radial and angular potentials, respectively. The subscript ``$n$'' denotes the near region, and ``$\pm_r$'', ``$\pm_{\t}$'' indicate outward and inward directions of $p^r$ and $p^{\t}$, respectively. It is obvious that the potentials must be non-negative. By demanding $R_n=R_n'=0$ , we obtain the critical impact parameters:
\bea\label{critical}
\td{\lm} &=& \f{a^2(M+\td{r})+\td{r}[2b+\td{r}(\td{r}-3M)]}{a(M-\td{r})} \ ,\nn\\
\td{\eta} &=&-\td{r}^2\f{4a^2(b-M\td{r})+[2b+\td{r}(\td{r}-3M)]^2}{a^2(M-\td{r})^2} \ .
\eea
This represents a critical curve on $(\lm, \eta)$-plane parameterized by $\td{r}$. The photons with such impact parameters move in spherical orbits of radius $r = \td{r}$, and all the spherical photon orbits make a spherical shell, which is called a photon region. 

\begin{figure}[H]
\centering
\includegraphics[scale=0.2]{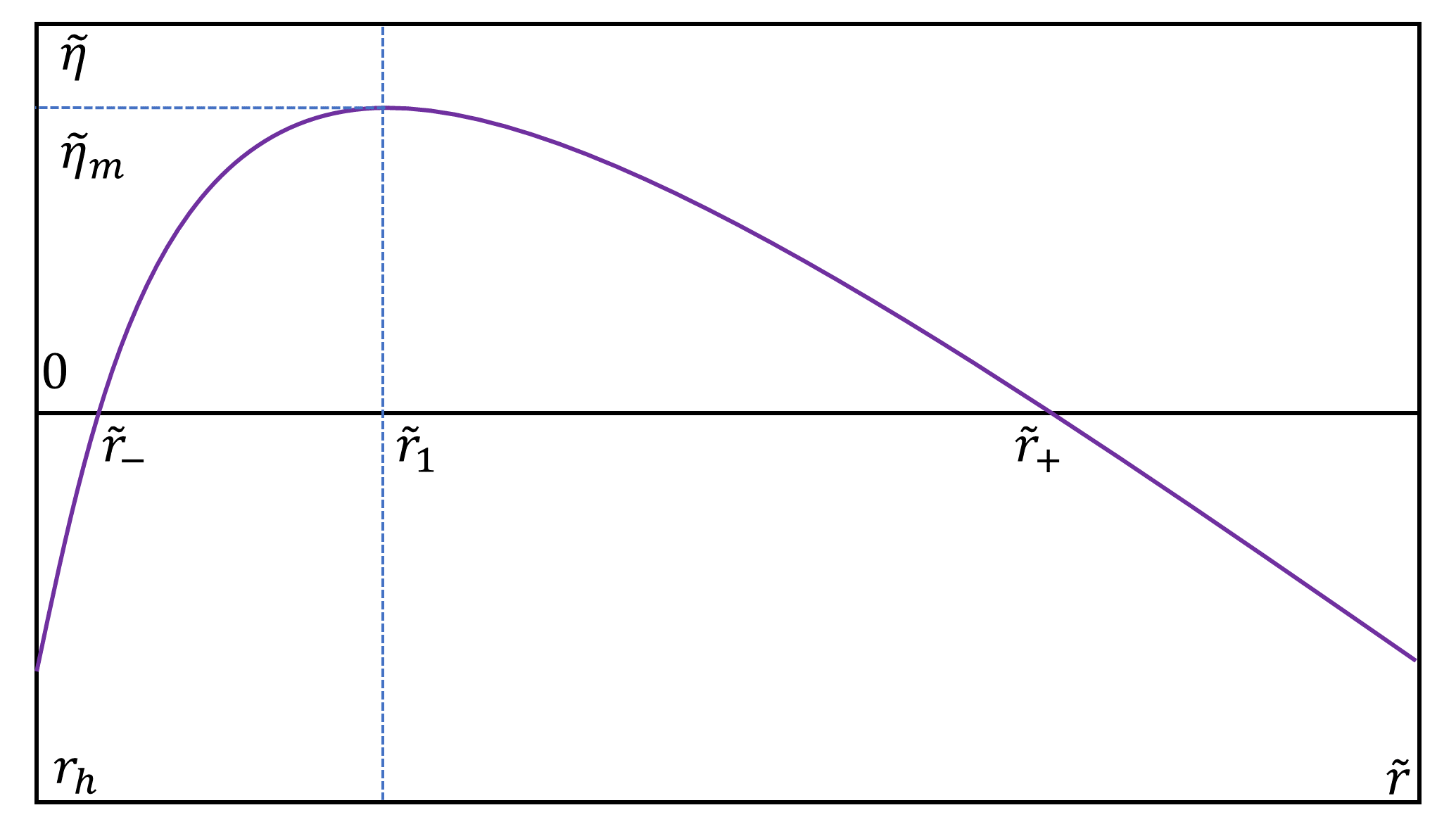}
\caption{The critical Carter constant $\td{\eta}(\td{r})$. The event horizon is at $r_h$, the bound is between $(\td{r}_- ,\td{r}_+)$, and the maximal value of $\td{\eta}$ is $\td{\eta}_m$ at $r = \td{r}_1$.}\label{etaex}
\end{figure}

Next, we would like to determine the allowed range of $(\td{\lm}, \td{\eta})$ of the photon region by requiring the angular potential to be non-negative. From the Eq. (\ref{Tn}), we find the angular potential $\T_n(\t)$ has the same form with the one in the Kerr solution. Thus, as in the Kerr case,  the photons that can pass through the equatorial plane must have non-negative Carter constant, that is $\eta\ge0$.  The photon region with $\eta \geq 0$ sometimes is referred to as a normal region \cite{Gralla:2019ceu}.
To verify the existence of such region outside the horizon, we need to know the behavior of the function $\td{\eta} (r)$. In the following, for simplicity and without loss of generality we set $M=1$. Firstly, we find that 
\bea
\td{\eta} (r_h)  = -\frac{(a^2+b-2r_h)^2}{a^2}< 0 \ ,
\eea
is always true. On the other hand, we conclude $\td{\eta} (r\rightarrow\infty) \sim -r^4\to-\infty$. In addition, from $\partial_{r}\td{\eta}$, we find that there is only one root 
\be
r_1 = \f1{2}\l( 3 + \sqrt{9-8b} \r) \ ,
\ee
outside the black hole horizon. Moreover, we check that $\partial_{r_1}^2\td{\eta} < 0$. Thus the function $\td{\eta}$ has a maximum,
\bea
\td{\eta}_m=\td{\eta}(r_1)=2\l(\f{ 3+\sqrt{9-8b} }{ 1+\sqrt{9-8b} }\r)^2 \l(3-2b+\sqrt{9-8b}\r)>0 \ ,
\eea
where we have used the fact that $b\le1$ in order to avoid naked singularity. Therefore, we conclude that the braneworld black holes with tidal charge $b\le1 - a^2$ always have a normal region $r_h<\td{r}_- <\td{r}< \td{r}_+$ outside the horizon, as illustrated in Fig. \ref{etaex}. The end points of the normal region satisfy $\td{\eta}=0$, which is a quartic equation and  has no analytical solution in general.  In addition, It is worth mentioning that $\td{r}=\td{r}_\pm$ correspond to the light ring on the  equatorial plane.

Next, we move to the angular potential $\T_n$ to figure out the photon region on the $(\lambda, \eta)$-space. One of the most important things is to find the minimum of $\td{\eta}(\td{\lambda})$ in the photon region. For this purpose, we start with a general analysis of the allowed $(\lambda, \eta)$ with $\T_n\ge0$. Let us introduce $x=\cos^2{\t} \in (0,1)$, and multiply both sides of the Eq. (\ref{Tn}) by $\sin^2{\t}$, then we have 
\be
(1-x)\T_n =f(x)= -a^2x^2 + (a^2-\lm^2-\eta)x + \eta \ .
\ee
The allowed photons are confined in the region that $f(x) \geq 0$. And the boundaries are located at the positions which satisfy $f(x)=0$. They  are 
\bea
x_{\pm}=X \pm \sqrt{X^2 +\f{\eta}{a^2}} \ , \ X = \f1{2}\l( 1-\f{\eta+\lm^2}{a^2} \r) \ .\label{root}
\eea
Thus, the roots of $\T_n(\t)=0$ are $\t_{\pm\pm} = \arccos{\pm\sqrt{x_\pm}}$. Obviously, to have a real $x_\pm$, the terms in the square root of Eq. (\ref{root}) must be non-negative, that is, $ a^2X^2+\eta \geq 0$, which gives us  
\bea\label{real}
\eta \geq -(|\lm|-a)^2, \hspace{3ex} \text{or} \hspace{3ex}  \eta \leq -(|\lm|+a)^2 \ .
\eea

The function $f(x)$ is a quadratic function of $x$ and its shape is a parabola pointing downwards. Considering the fact that $f(0) = \eta$ and $f(1) = -\lm^2 \leq 0$, the number of roots of $f(x)=0$ is dependent on the sign of $\eta$. When $\eta > 0$, the function $f$ definitely has one zero $x_+$ in $(0,1)$. In this case, $x_+$ is always positive and $x_-$ is always negative, thus the allowed region in the angular direction is $\theta_{++}\le\theta\le\theta_{-+}$. Since $\theta_{++}<\pi/2$ and $\theta_{-+}>\pi/2$, the photons in this region must pass through the equatorial plane repeatedly. When $\eta = 0$, we find $x_+=0$ and $x_-<0$ for $|\lambda|>a$, while we have $x_-=0$ and $0<x_+=2X\leq1$ for $|\lambda|<a$. In particular, $|\lambda|=a$ gives us $x_\pm=0$. That is to say, no matter which is bigger between $|\lambda|$ and $a$, the null geodesics with $\eta=0$ always exist, and in particular,  the photons with $\eta=0$ are confined on the equatorial plane when $|\lambda|\ge a$.

Next, we turn to the region $\eta < 0$. The function $f(x)$ would have two or no zeros in $x\in (0,1)$. Under the condition Eq. (\ref{real}), to ensure a region $f(x) \geq 0$ in $(0,1)$, the axis of symmetry of $f(x)$ must be in $(0,1)$, that is, $0<(a^2-\lm^2-\eta)/2a^2<1$, which gives
\be\label{max}
-a^2-\lm^2 < \eta < a^2 - \lm^2 \ .
\ee
Combining Eq.(\ref{real}) with Eq.(\ref{max}), we find that the allowed region is $ 0 > \eta \geq -(|\lm|-a)^2$ for $|\lm|\leq a$. In this situation, $f(x)$ has two zeros $x_+$, $x_-$ in $(0,1)$. This is the so called vortical region, admitting four real roots $\t_{++}  < \t_{+-}  < \f{\pi}{2} < \t_{--} <\t_{-+}$, with the potential being positive between two intervals: $(\t_{++} , \t_{+-})$ and $(\t_{--} , \t_{-+})$, so the photons librate between one of the two intervals, never crossing the equatorial plane.

\begin{figure}[H]
\centering
\includegraphics[scale=0.55]{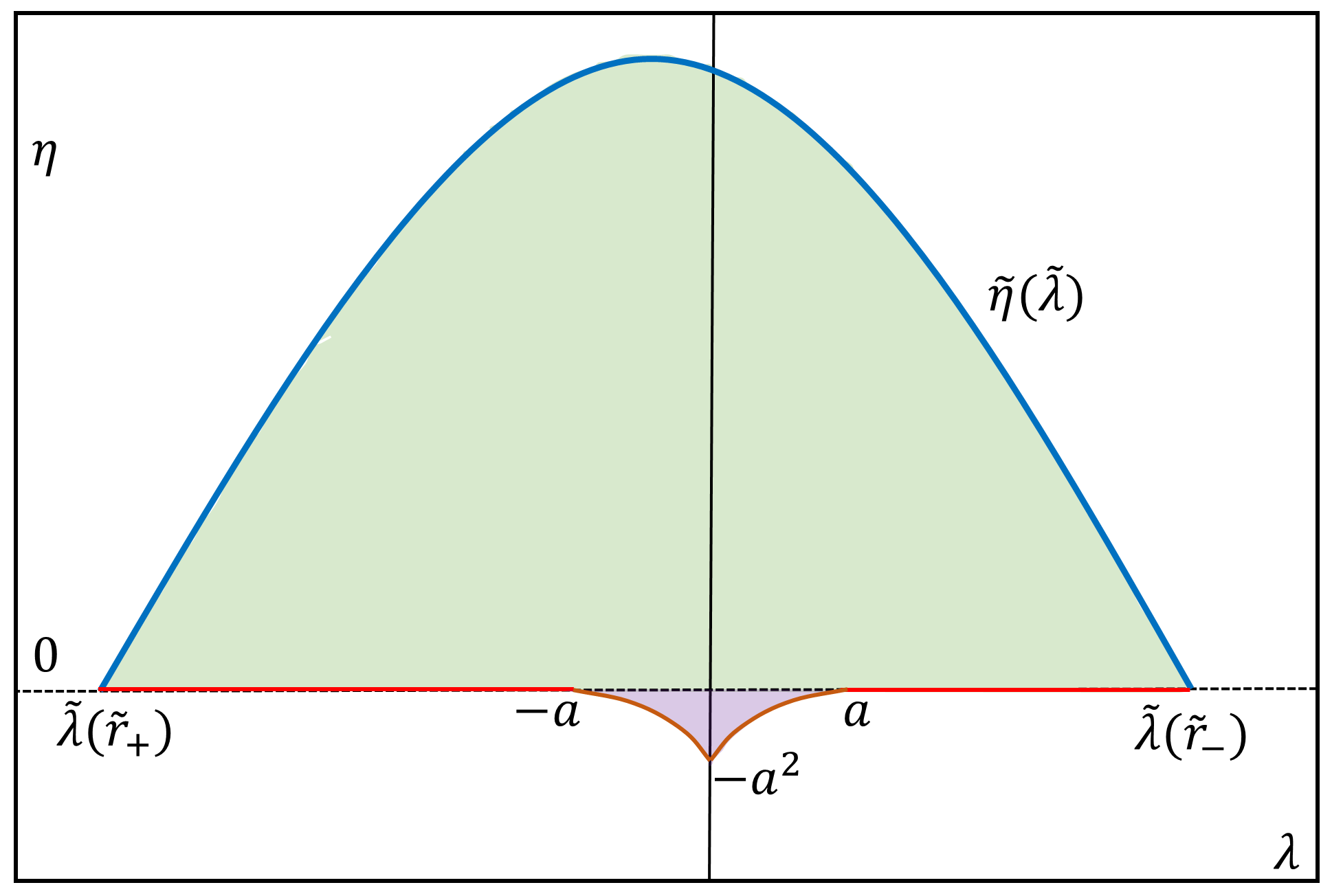}
\caption{The blue curve denotes the critical curve $\td{\eta}(\td{\lm})$, whose left and right end points correspond to $\td{r}_+$ and $\td{r}_-$ respectively. The green zone is normal region, and the grey zone is vortical region. }\label{ep}
\end{figure}

In conclusion, we obtain the following constraints from the angular potential $\T(\t)$, which is the same as in \cite{Gralla:2019ceu}
\be\label{bound1}
\eta \geq \left\{
\ba{ll}
0 \ ,     &|\lm|\geq a \ , \\
-(|\lm|-a)^2,    &|\lm|\leq a \ .
\ea
\right. 
\ee
Now we are ready to deal with the allowed region of   $(\td{\lambda}, \td{\eta})$. Here, we would like to reiterate that the regions above and below the $\lm$-axis are the normal and vortical regions, respectively. In fact, from the Eqs.(\ref{critical}), one can find the function $\td{\eta}(\td{\lambda})$ in principle. Combining with the Eq.(\ref{bound1}), we are very near to have a clear idea of the photon region in the $(\lambda, \eta)$-plane. There remains only one question unclear, that is, we have no idea which is bigger between $a$ and $\td{\lambda}(\td{\eta}=0)$. To solve this problem, one way is to diagnose the sign of $\td{\eta}(\td{\lambda}=\pm a)$. For $\td{\lm} = a$, we find  
\be
\bar{r}_+\equiv r(\td{\lm} = a) = \f1{2}\l[ 3 + \sqrt{9-8(a^2+b)} \r] > r_h \ , 
\ee
and then  $\td{\eta}(\bar{r}_+) > 0$. As for $\td{\lm}(\bar{r}_-) = -a$, the expression of $\bar{r}$ is very complicated, thus we omit the exact expression here. After some algebraic manipulations, we find  
\be
\td{\eta}(\td{\lm} = -a) = \f{4}{(1-r)^2}\l[ r^2(r-b)-a^2 \r]|_{r=\bar{r}_-} \ . 
\ee
at $\td{\lm} = -a$. Considering $\bar{r}_-$ is larger than $r_h$, it is easy to see that $\l[ r^2(r-b)-a^2 \r]|_{r=\bar{r}_-}  > (1-b)-a^2 \ge 0$. Therefore, we can conclude $\td{\eta}(\td{\lm} = \pm a) >0 $. To end up, we find the critical curve is parameterized by the radius $\td{r}$ in the photon region $(\td{r}_-,\td{r}_+)$, with the right endpoint $(\td{\lm}(\td{r}_-),0)$ and the left endpoint $(\td{\lm}(\td{r}_+),0)$. Also, we find $|\td{\lambda}(\bar{r}_\pm)|>a$, thus for any $a$ and $b$, all the photon regions on the $(\td{\lambda}, \td{\eta})$ plane are in the shape shown in Fig. \ref{ep}  

\subsection{Null geodesics in the far region}
In this section, we turn to the far region, that is $r\gg r_h$. In this region, the spacetime can be described by the linearized metric \cite{RWW}
\bea\label{fs}
ds^2 =& -&\l[ 1 - \f{2M}{r}\l( 1+ \f{2l^2}{3r^2} \r) \r]dt^2 \ + \ \l[ 1 - \f{2M}{r}\l( 1+ \f{l^2}{r^2} \r) \r]^{-1}dr^2 \ + \ r^2 d\t^2 \nn  \\
&+&\ r^2 \sin^2{\t}  \l[ d \p - \f{2Ma}{r^3} \l( 1+ \f{3l^2}{2r^2} \r)dt \r]^2 \ ,
\eea
where $l$ is the curvature length of the $\text{AdS}_5$ spacetime in the bulk. 
From this metric, we can see that it may describe a slow-rotating spacetime in the weak field approximation if $l$ vanishes. In other words, if $l$ is absent and we drop the terms that contains $a^2$,  this metric could be reproduced from the metric (\ref{ns}) in the large $r$ limit.  Notice that  the linearized metric cannot be extended to the near region either. The fact that two metrics \eqref{ns} and \eqref{fs} describe two different spacetime regions without overlapping can be seen from the mismatch of the radial motions of null geodesics in two regions.

Similarly, we can separate the variables in Hamilton-Jacob equation of null geodesics in the far region
\bea
p_{\t}^2 +  \f{\lm^2}{\sin^2{\t}} = -[r^2-2(\f{l^2}{r^2}-1)r]p_r^2 + \f{3}{r^3} \f{ \l[ r^3-a\lm(2+3l^2/r^2)\r]^2 }{3r-2(3+2l^2/r^2)}
 = K_f \ ,
\eea
where $K_f $ is a constant from separating the variables, and the subscript ``$f$ '' denotes the far region. We assume that the coordinates $( t , r , \t ,\p )$ cover the entire spacetime. In other words,  the metric (\ref{ns}) in the near region and the metric (\ref{fs}) in the far region can be regarded as two different limits of an unknown metric which cover the entire spacetime. Consequently,  the conserved quantities $(E, L, K)$, which are determined by the symmetries of spacetime, are valid throughout the whole spacetime, from the near region to the far region. Therefore, we can match $K_f$ with $K_n$, that is, 
\bea
K_f  = K_n = K = \eta + (\lm -a)^2 \ .
\eea
Then, the equations of motions along $r$ and $\t$ are
\bea\label{feqm}
r^2\dot{r} = \pm_r \sqrt{R_f}, \hspace{3ex} \ r^2\dot{\t} = \pm_{\t} \sqrt{\T_f} \ ,
\eea
where $R_f(r)$ and $\T_f(\t)$ are the potentials in the far region
\bea
R_f(r) &=& r^4\l[1+\f{2(l^2-r^2)}{r^3}\r]\l[\f{(1-A_2)^2}{A_1} -B \r] \ ,\\
\T_f(\t) &=&  \eta + (\lm -a)^2 - \f{\lm^2}{\sin^2{\t}} \ ,
\eea
and for simplicity we have defined 
\bea\label{AA}
A_1\equiv 1-\f{2}{r}(1+\f{2l^2}{3r^2}) \ , \ A_2\equiv \f{a\lm}{r^3}(2+3\f{l^2}{r^2}) \ ,
\eea
and 
\bea\label{BB}
B \equiv (\eta + (\lm -a)^2)/r^2.
\eea
In the far region, a non-negative potential is also required. For the radial direction, we can see that  the radial potential $R_f(r) \sim r^4$ is always positive when $r \gg 1$, while the requirement of a non-negative angular potential gives another constraint on $(\lm, \eta)$
\be\label{bound2}
\eta \ \geq \ \lm^2\cot^2{\t} + 2a\lm - a^2\ge2a\lambda-a^2 \ .
\ee
note that the parameter $a$ is the spin of braneworld black hole and the same as the one in the metric of the near region.
\begin{figure}[H]
\centering
\includegraphics[scale=0.3]{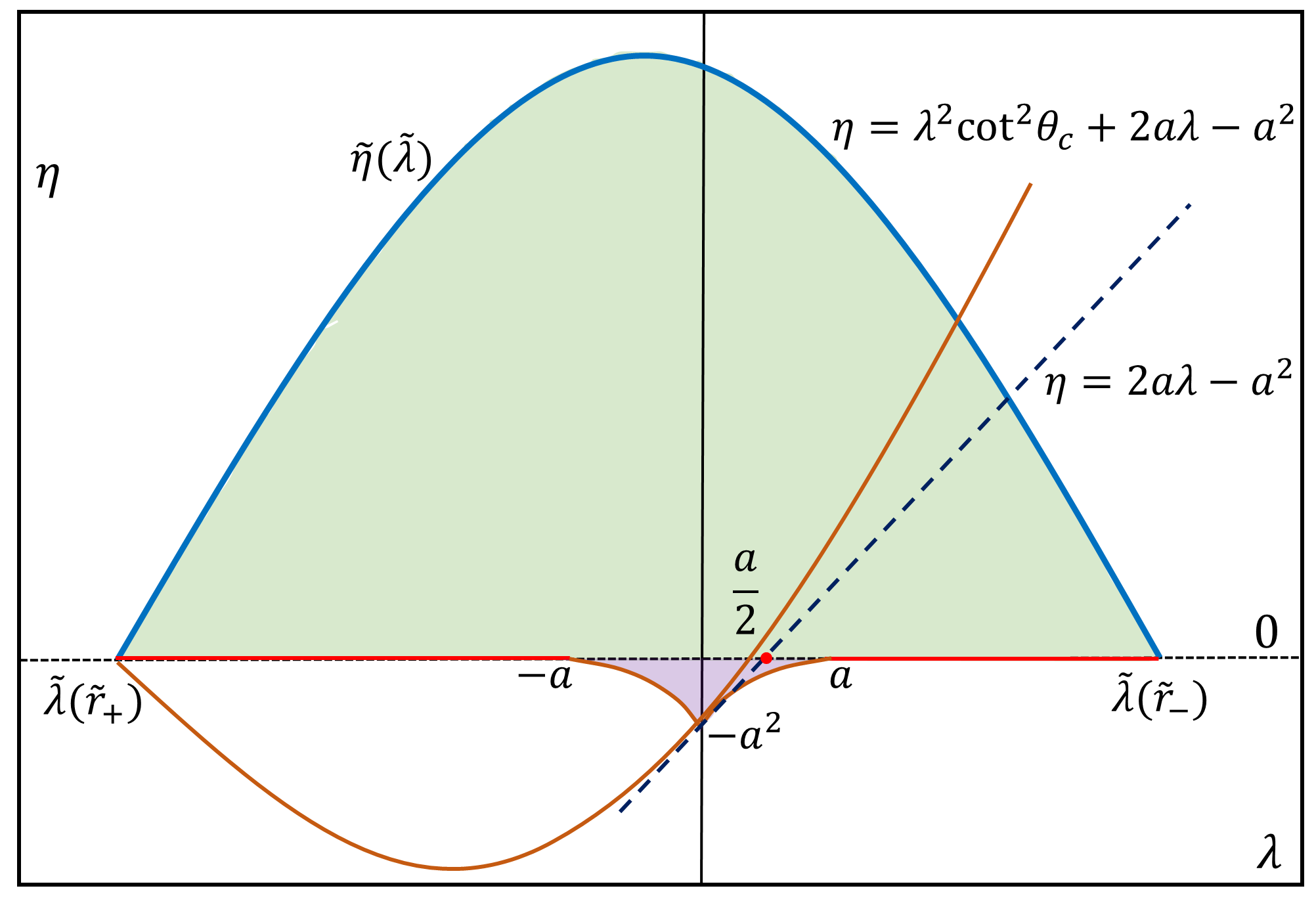}\\
\caption{The red curve is the lower bound of \eqref{bound2} with a critical angle.}\label{C}
\end{figure}

\begin{figure}[H]
\centering
\includegraphics[scale=0.36]{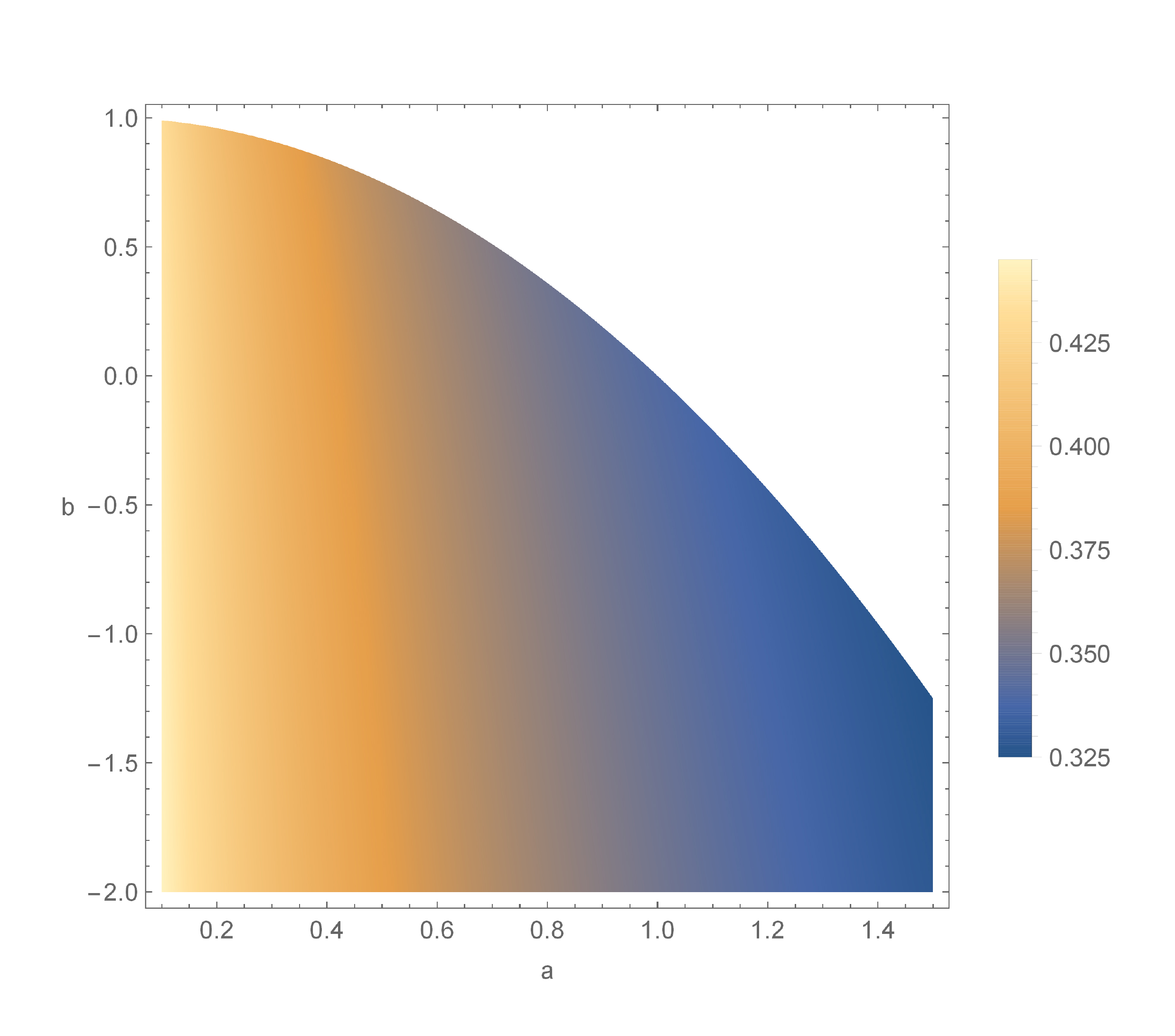}
\caption{Density plot of critical angle $\t_c/\pi$ as a function of $a$ and $b$. The ranges are $a \in (0.1,1.5)$, $b \in (-2 \ ,1-a^2)$. }\label{tc1}
\end{figure}

\begin{figure}[H]
\centering
\includegraphics[scale=0.35]{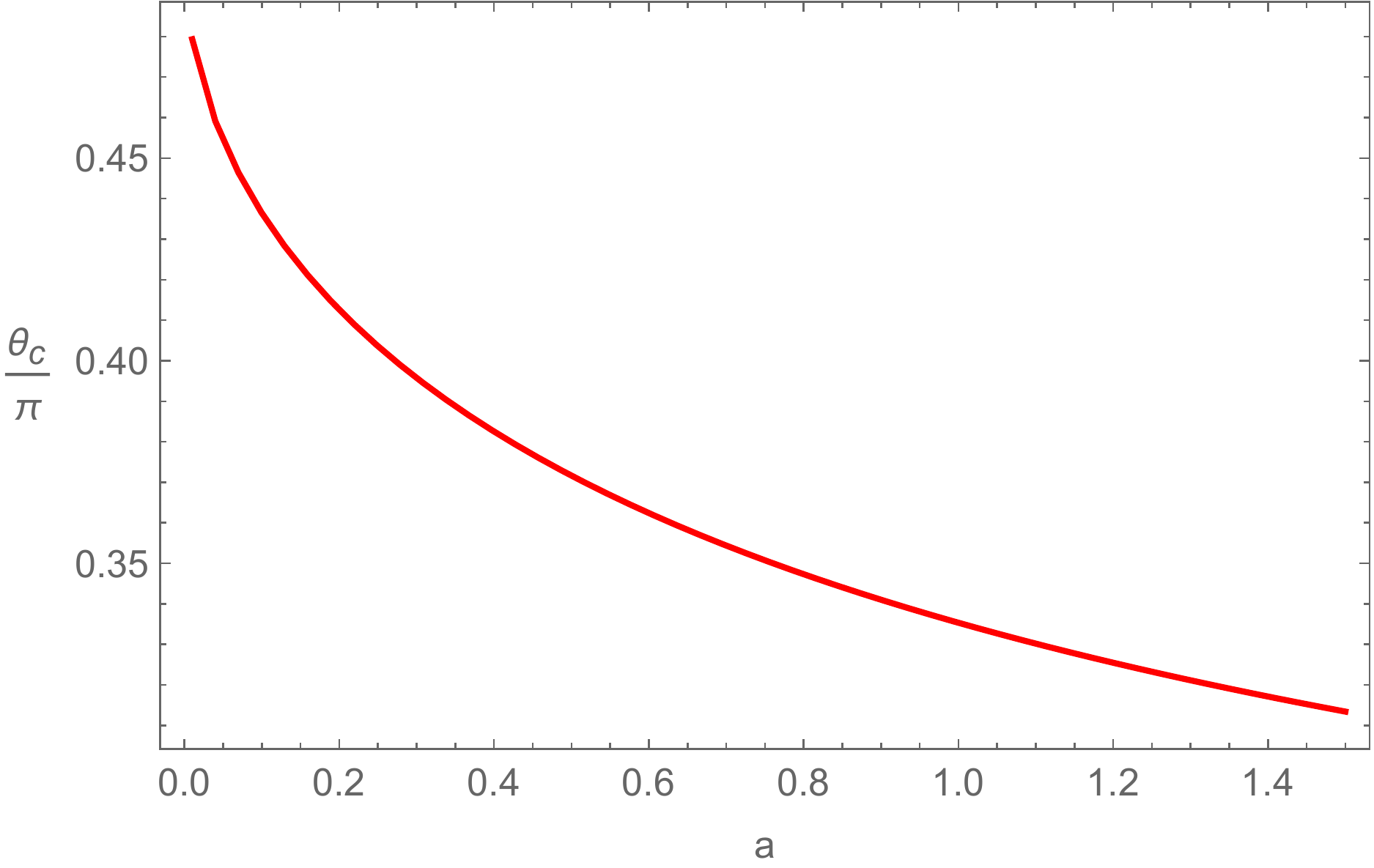}  \quad \includegraphics[scale=0.35]{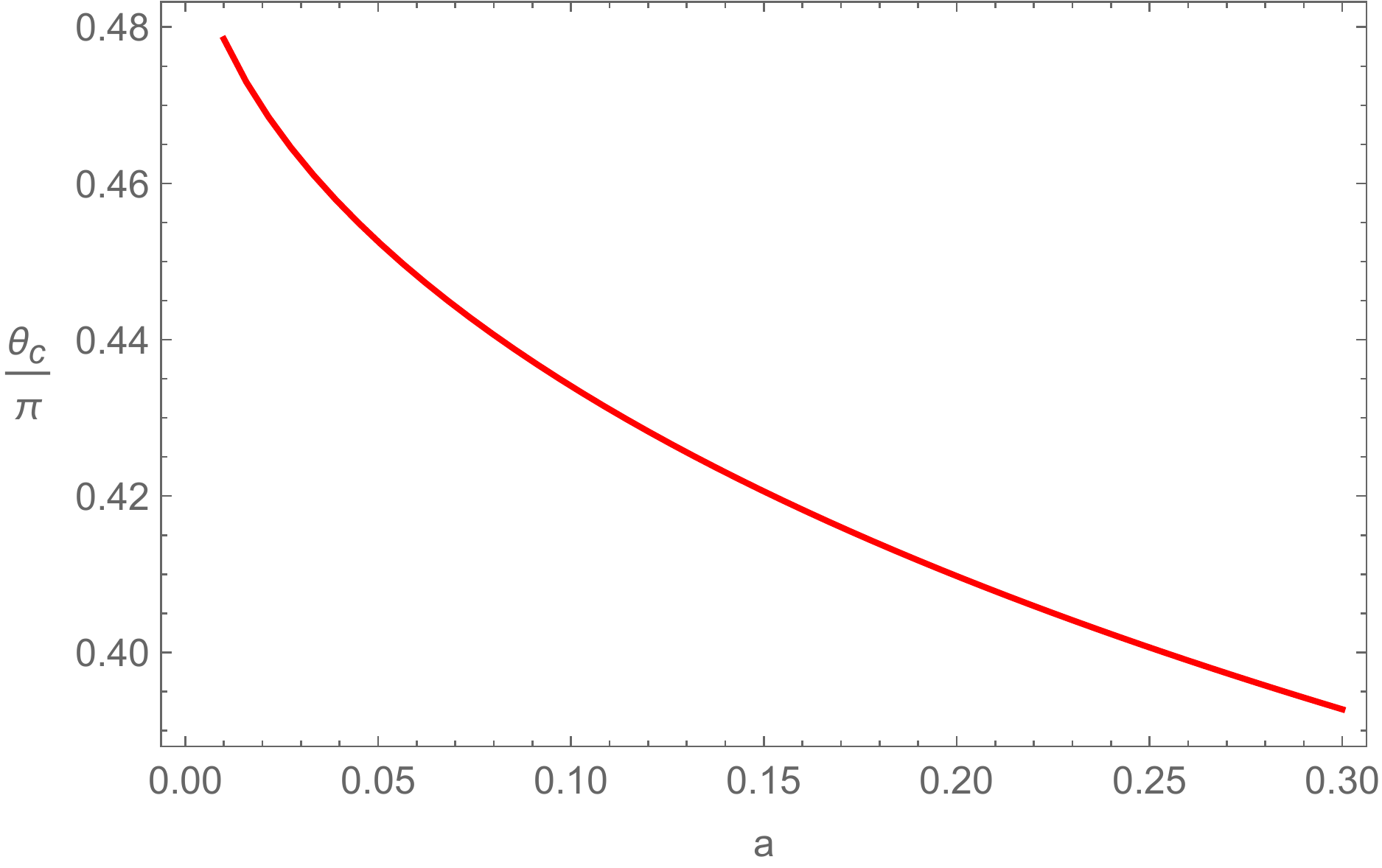} \\
~\\
\includegraphics[scale=0.35]{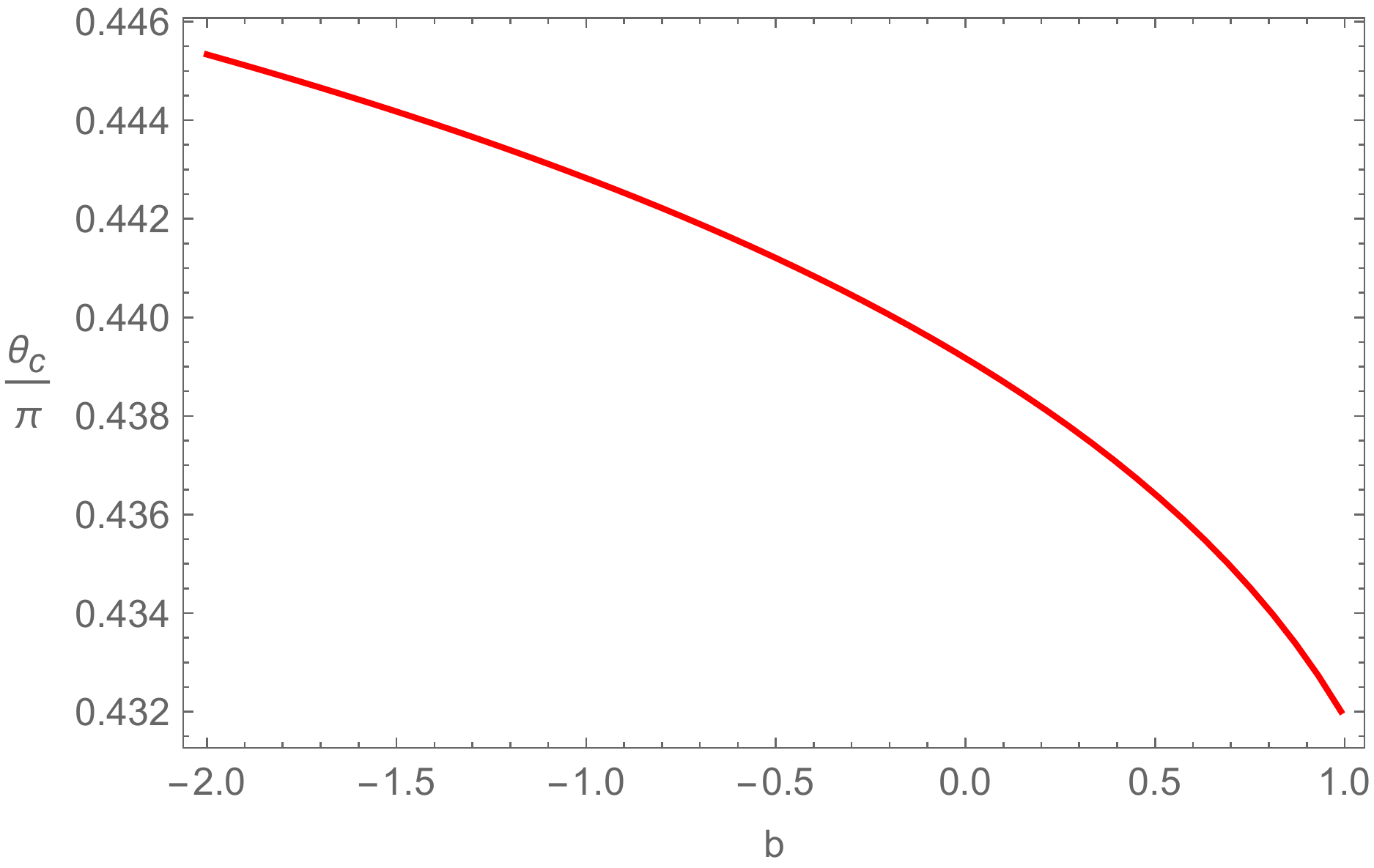}  \quad \includegraphics[scale=0.35]{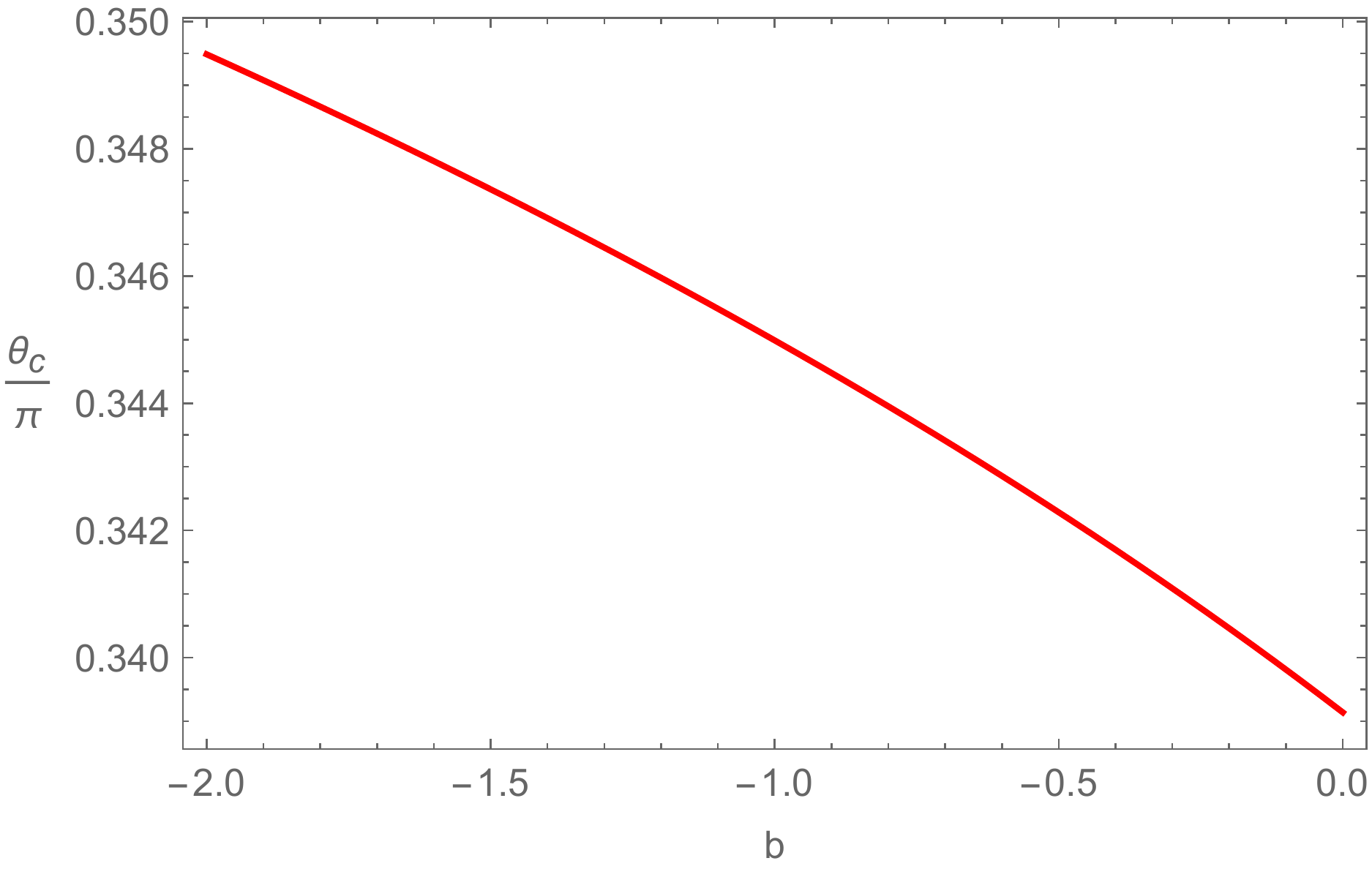}\\
\caption{Top left: $\t_c$ changes with $a$, the tidal charge is fixed to be $b=-2$. Top right: $\t_c$ changes with $a$, and  $b=0.8$. Down left: $\t_c$ changes with $b$, with $a = 0.1$. Down right:  $\t_c$ changes with $b$, and $a =1$.}\label{tc2}
\end{figure}

As seen in Fig. \ref{C}, the critical curve $\td{\eta}(\td{\lambda})$ and the lower bound $\eta=2a\lambda-a^2$ has a intersection point since the lower bound goes through $(0, -a^2)$ and $(a/2, 0)$. When $\theta<\pi/2$, the lower bound $\eta(\lambda)$ is always a parabola going upwards and above the line $\eta=2a\lambda-a^2$. Thus, the critical curve $\td{\eta}(\td{\lambda})$ and the lower bound $\eta(\lambda)$ at least has an intersection point for any $\theta\in(0, \pi)$. Notice that there exists a critical angle $\theta_c$ that $\eta(\lambda)$ would go cross the point $(\lm(\td{r}_+),0)$, where $\td{r}_+$ is the larger root of $\td{\eta}=0$. This leads to the critical angle 
\be
\t_c(a,b) = \arcsin{\l\{\f{\lm(\td{r}_+)}{\lm(\td{r}_+)-a}\r\}} \ .
\ee
Considering the $\mathcal{Z}_2$ symmetry, we can focus on $0<\theta\le\pi/2$. Furthermore we find that $\td{\eta}(\td{\lambda})$ and $\eta(\lambda)$ would have two intersections when $0<\theta\le\theta_c$ and only a single intersection when $\theta_c<\theta\le\pi/2$. This interesting finding would have an important role in determining the shadow curve of braneworld black holes, as we will discuss in detail  in section \ref{shadow}.

Next, we want to examine the variation of the critical angle $\theta_c$ with respect to $a$ and $b$. Recall that a braneworld black hole with an event horizon allows a negative $b$, thus the spin $a$ can be larger than the mass of the black hole. We expect that the spin would not be too large, thus we choose $a$ goes from $0.1$ to $1.5$ in this work, and we let $b$ range from $-2$ to $1-a^2$ to respect the weak cosmic censorship conjecture. Numerically, we obtain the density plot of $\theta_c/\pi$ with respect to $a$ and $b$ in Fig, \ref{tc1}. From this figure, we find that when a braneworld black hole has a larger spin and tidal charge, the critical angle would be smaller. In the range of $a$ and $b$ that we are thinking about, the minimum value $\theta_c^{min}=0.325\pi$ which means if $\theta\le\theta_c^{min}$, the critical curve $\td{\eta}(\td{\lambda})$ and the lower bound $\eta(\lambda)$ always have two intersection points. This important fact  will be used in the later discussions.

In addition, in Fig. \ref{tc2} we show the variation of the critical angle $\theta_c$ with respect to $a$ or $b$  by fixing the other parameter. For a fixed $a$, we examine a low spin case $a=0.1$ and a high spin case $a=1$. For a fixed $b$, both a positive example $b=0.8$ and a negative example $b=-2$ are displayed. We find that for all of these cases, $\theta_c$ are monotonically decreasing.

\section{Shadows in the sky of distant observers}\label{shadow}

In this section we would like to investigate the shadows viewd by distant observers. Considering a static observer at $ (t_o=0, r_o ,  \t_o , \p_o = 0)$ in the far regon, with its basis vectors \cite{Bardeen:1973tla}
\bea
\hat{e}_{(t)}&=&\sqrt{\frac{g_{\phi\phi}}{g_{t\phi}^2-g_{tt}g_{\phi\phi}}}\left(\partial_t-\frac{g_{t\phi}}{g_{\phi\phi}}\partial_\phi\right),\nn\\
\hat{e}_{(r)}&=&\frac{1}{\sqrt{g_{rr}}}\partial_r,\nn\\
\hat{e}_{(\theta)}&=&\frac{1}{\sqrt{g_{\theta\theta}}}\partial_\theta,\nn\\
\hat{e}_{(\phi)}&=&\frac{1}{\sqrt{g_{\phi\phi}}}\partial_\phi,
\eea
where the metric components are given in Eq. (\ref{fs}), we obtain the projections of 4-momentum in the frame of the observer
\be\label{basis}
\bag
p^{(i)}&=\hat{e}_{(i)}^{\m} p_{\m}=\f{1}{\sqrt{g_{ii}}} p_{i} \ , \ i=r,\t,\p \ ,  \\
p^{(t)}&=E \z-L\g \ .
\eag
\ee
Due to the drag effect of the rotating black hole, $p^{(t)}$ is the mixture of the energy $E$ and the angular momentum $L$ with
\bea
\z=\sqrt{\f{g_{\p\p}}{g_{t \p}^{2}-g_{t t} g_{\p\p}}} \ , \hspace{3ex} 
\g=-\f{g_{t\p}}{g_{\p\p}} \sqrt{\frac{g_{\p\p}}{g_{t \p}^{2}-g_{t t} g_{\p\p}}} \ .
\eea
On the other hand, we may introduce the observational angels $(\alpha, \beta)$ in the observer's local rest frame such that 
\bea
p^{(r)}&=&|\vec{P}|\cos\alpha\cos\beta,\nn\\
p^{(\theta)}&=&|\vec{P}|\sin\beta,\nn\\
p^{(\phi)}&=&|\vec{P}|\cos\beta\sin\alpha,
\eea
then, we have approximatively  in the far region $r_o\gg 1$ 
\be
\a = -r_o\frac{p^{(\p)}}{p^{(r)}} \ , \hspace{3ex}  \b = r_o\frac{p^{(\t)}}{p^{(t)}} \ .
\ee
Using Eq.(\ref{basis}) and Eq.(\ref{feqm}), we obtain the relations between the impact parameters $(\lm, \eta)$ and the coordinates of the photons in the sky of the observer, 
\bea
\a &=& -\f{\lm}{\sin{\t_o}} \ \a_l \label{af} \ , \\
\b &=&   \pm_{\t} \sqrt{ \T_f(\t_o) } \ \b_l \label{bf} \ ,
\eea
where
\begin{equation}
\begin{aligned}
\a_l &=& \sqrt{\f{A_1}{A_2^2+A_1(6A_2-B^2)+1}} \ , \hspace{3ex}
\b_l &=& \f{\sqrt{A_1}}{1-A_2} \ ,
\end{aligned}
\end{equation}
with $A_1,A_2$ and $B$ being given in Eqs. (\ref{AA}) and (\ref{BB}) and valued at $r=r_o$. Eqs.(\ref{af}) and (\ref{bf}) can be seen as a map between $(\lm, \eta)$ and $(\a, \b)$. If the photons with impact parameters $(\lm, \eta)$ reach the observer, they will leave specific light spots in the image plane and the parameters should satisfy $\Theta_f(\theta_o) \geq 0$, which is just Eq. (\ref{bound2}). On the other hand, if we choose a light spot $(\a,\b)$ and trace the light back, we will find the geodesic carrying specific impact parameters.

\begin{figure}[H]
\centering
\includegraphics[scale=0.4]{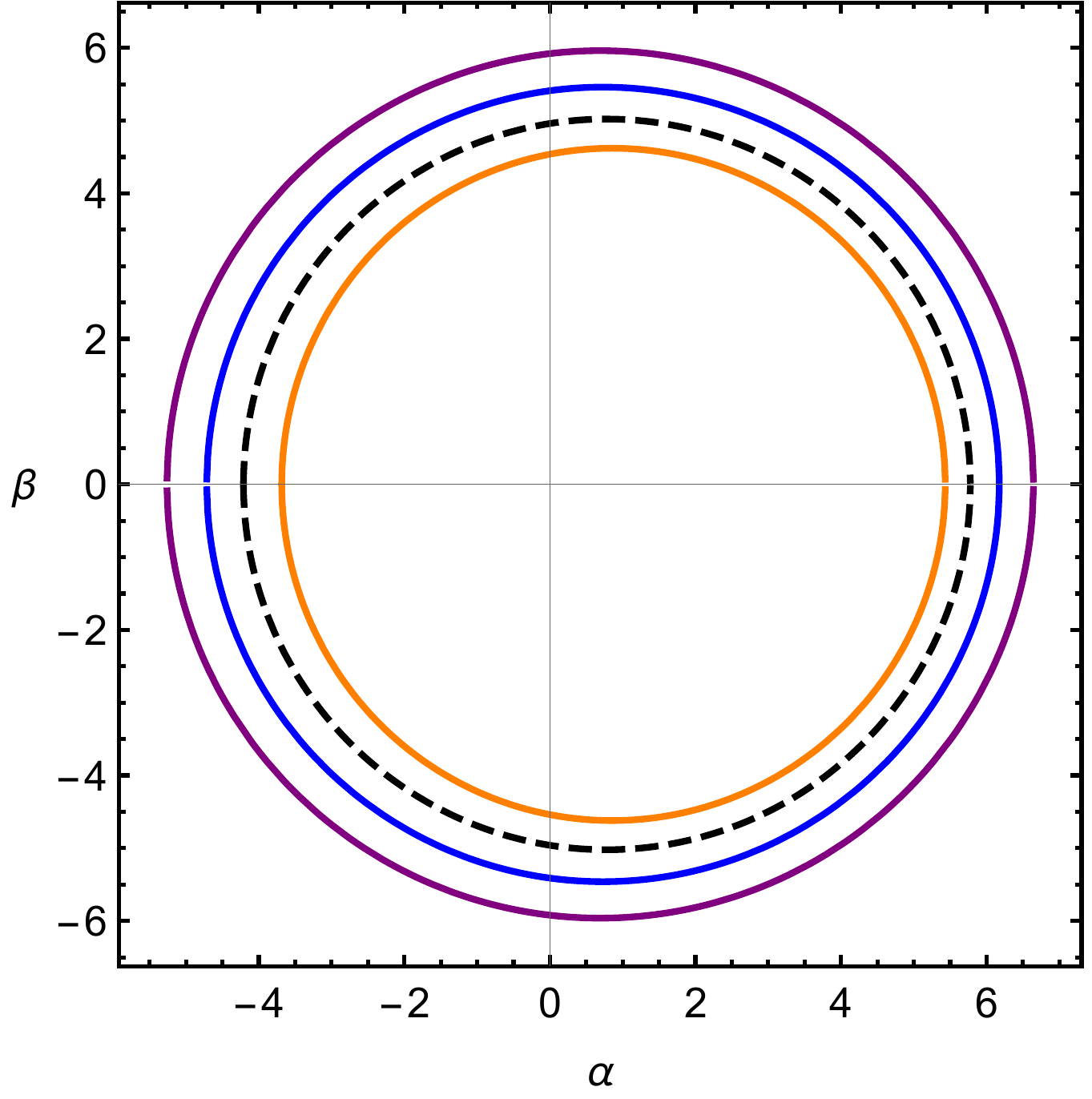} \  \includegraphics[scale=0.4]{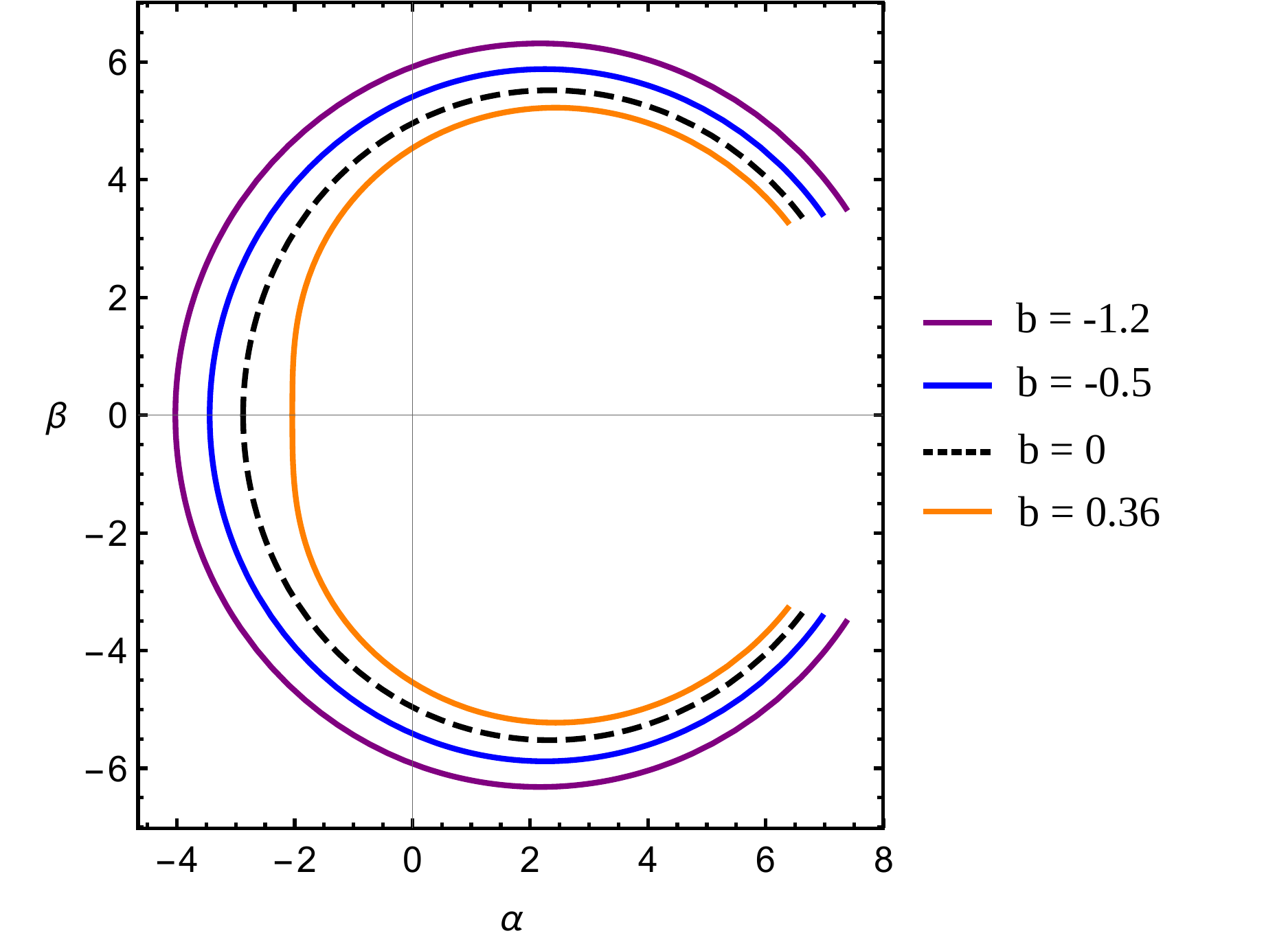} \\
\caption{Critical curves with different tidal charge $b$. We have set $M=1$,  $a=0.8$, $l=5$ ; and the azimuthal angle $\t_o = \f{\pi}{10}$(left) and $ \f{\pi}{2}$(right), the distance $r_o = 200$. }\label{C2}
\end{figure}

\begin{figure}[H]
\centering
\includegraphics[scale=0.42]{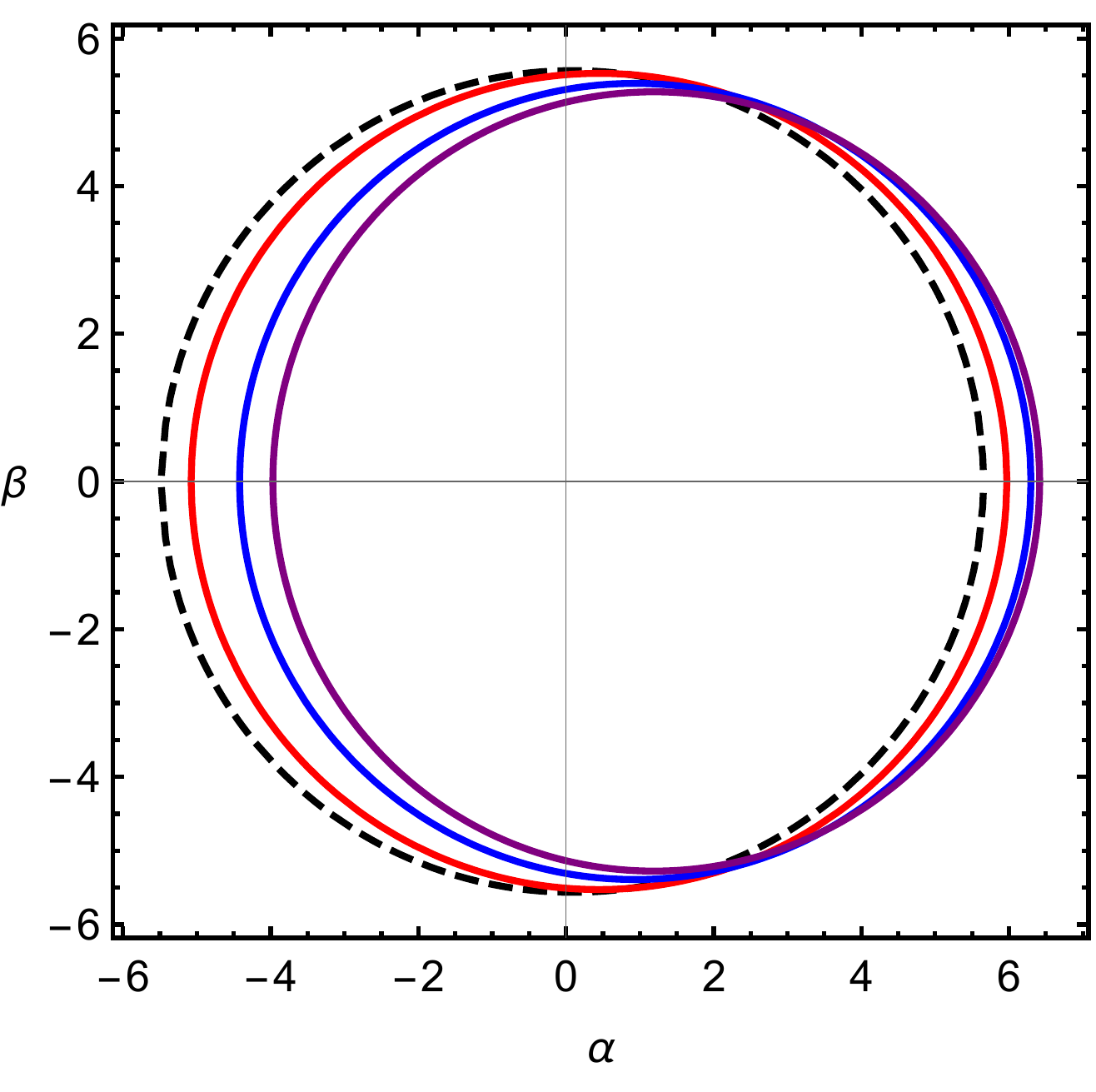} \quad  \includegraphics[scale=0.41]{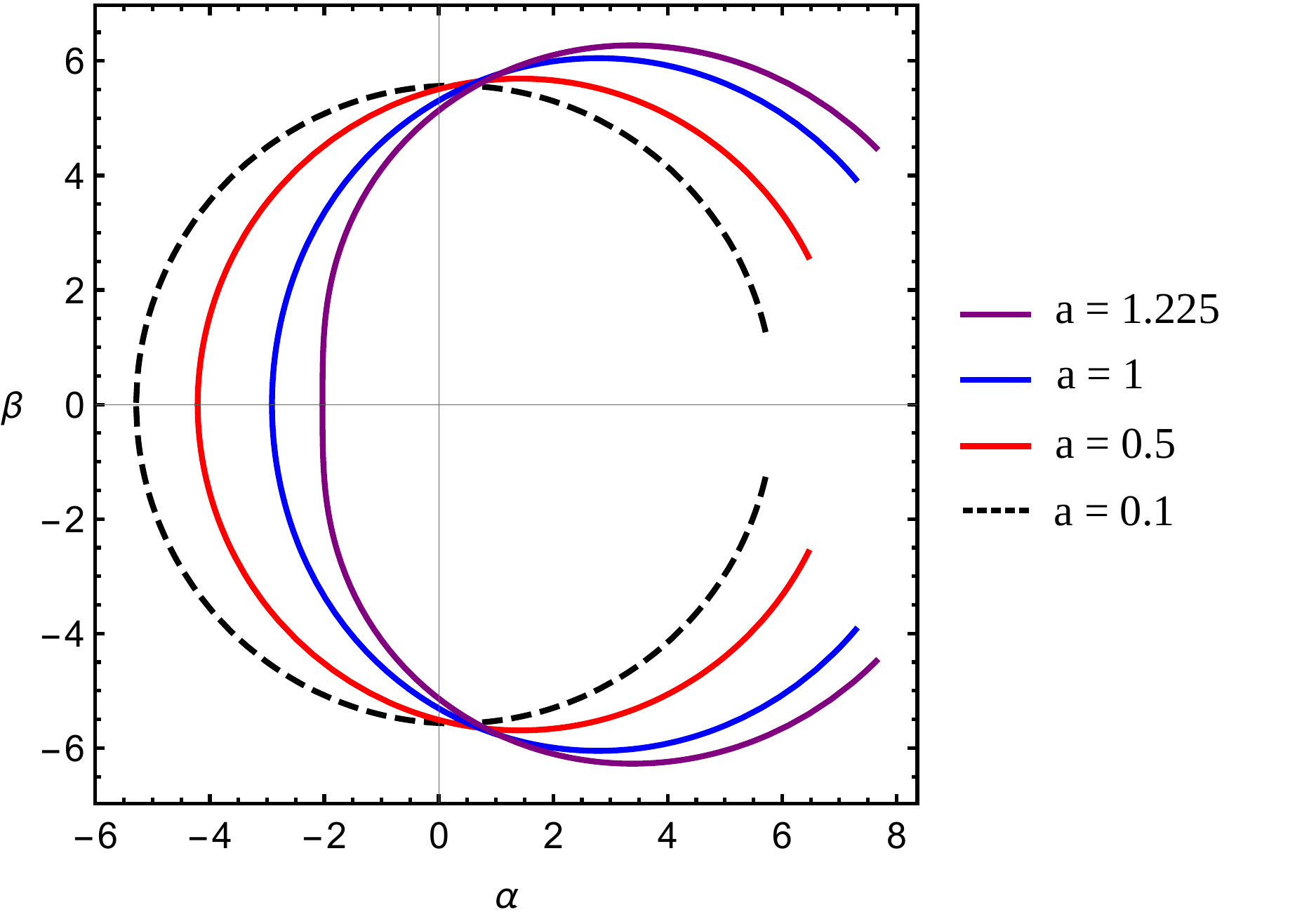} \\
\caption{Critical curves with different spin $a$. We have set $M=1$,  $b=-0.5$, $l=5$; and the azimuthal angle $\t_o = \f{\pi}{10}$(left) and $ \f{\pi}{2}$(right), the distance $r_o = 200$.}\label{C3}
\end{figure}

Combining Eqs. (\ref{af}), (\ref{bf}) with (\ref{critical}), we obtain the critical curve $C_i$ on the image plane, which is parameterized by the radius of the photon region $\td{r}$. Eq. (\ref{bf}) tell us that $C_i$ is symmetric about the $\a$ axis, as seen in Fig. \ref{C2} and Fig. \ref{C3}. Recall that the critical curve $\tilde{\eta}(\tilde{\lambda})$ and the bound curve $\eta(\lambda)$ can have one or two intersection points which corresponds to the fact that the shadow curve is open or closed. As we have discussed in the last section, there exist a critical angle to decide the intersection points and thus determine whether the shadow curve is closed or open. More precisely, the shadow curve would be closed when the observational $0<\theta_o\le\theta_c$ and the shadow curve would be open when $\theta_0>\theta_c$. In the present work,  we consider the black holes with the parameters  $a\in(0.1, 1.5)$ and $b\in(-2, 1-a^2)$, the minimum of $\theta_c$ is found to be $\theta_c^{min}=0.325\pi$ and the maximum of $\theta_c$ is identified as $\theta_c^{max}=0.44\pi$. Therefore  if the observational angle is less than $\theta_c^{min}$, the shadow curve is always closed and if the observational angle is between $\theta_c^{min}$ and $\theta_c^{max}$, whether the shadow curve is closed or open depends on the values of $a$ and $b$. And if the observational angle is bigger than $\theta_c^{max}$, the curve is always open. By the way, the curve in  \cite{Amarilla:2011fx} is also an unclosed curve, but it is open on the left due to the naked singularity.

In the following, we would like to study the influences of $a$, $b$, $l$ and $r_o$ on the shadow curves, respectively. Because of the possible existence of open shadow curves, we will show the results for two different angles, one of which would give a closed shadow curve and the other one gives an open curve. For the closed one, we choose $\theta_o=\pi/10<0.325\pi$ and we choose $\theta_o=\pi/2>0.44\pi$ to give the open one.

In Fig. \ref{C2}, we present the shadow curves with various tidal charges $b$. It can be seen that the size of the closed (open) curve becomes smaller with an increasing $b$ for $\theta_o=\pi/10$ ($\theta_o=\pi/2$), which is consistent with the result in \cite{BSS}. 

In Fig. \ref{C3}, we study the shadow curves with different spins, and we find that as the spin $a$ becomes bigger, the shadow curve is getting a little flatter. When $a$ goes near its upper limit, that is, Kerr black hole becomes near extremal, the left portion of the curve includes a vertical line segment, which is the so-called NHEK-line. The NHEK-line is named in \cite{Gralla:2017ufe} to denote the segment that the photons near the line originate from the near horizon region of extremal Kerr black hole (NHEK)  spacetime.  When Kerr black hole is near the extreme, the photons near the edge of the shadow curve are from the NHEK region.  More examples can be seen in \cite{Guo:2018kis, Yan:2019etp, Guo:2019lur}. Moreover, for the NHEK geometry, an enhanced symmetry arise which allow analytic calculations\footnote{Another feasible analytic study on the photon emissions in the rotating black holes is based on large $D$ expansion of gravity, see \cite{Guo:2019pte}.} in many cases including the null geodesics \cite{Porfyriadis:2016gwb, Li:2020val}. 

In addition, for $\theta_o=\pi/2$, since the end point of the curve is identified with $\tilde{r}_+$ which is sensitive to the spin $a$, the end points of the shadow curves are obviously different. 

\begin{figure}[H]
\centering
\includegraphics[scale=0.4]{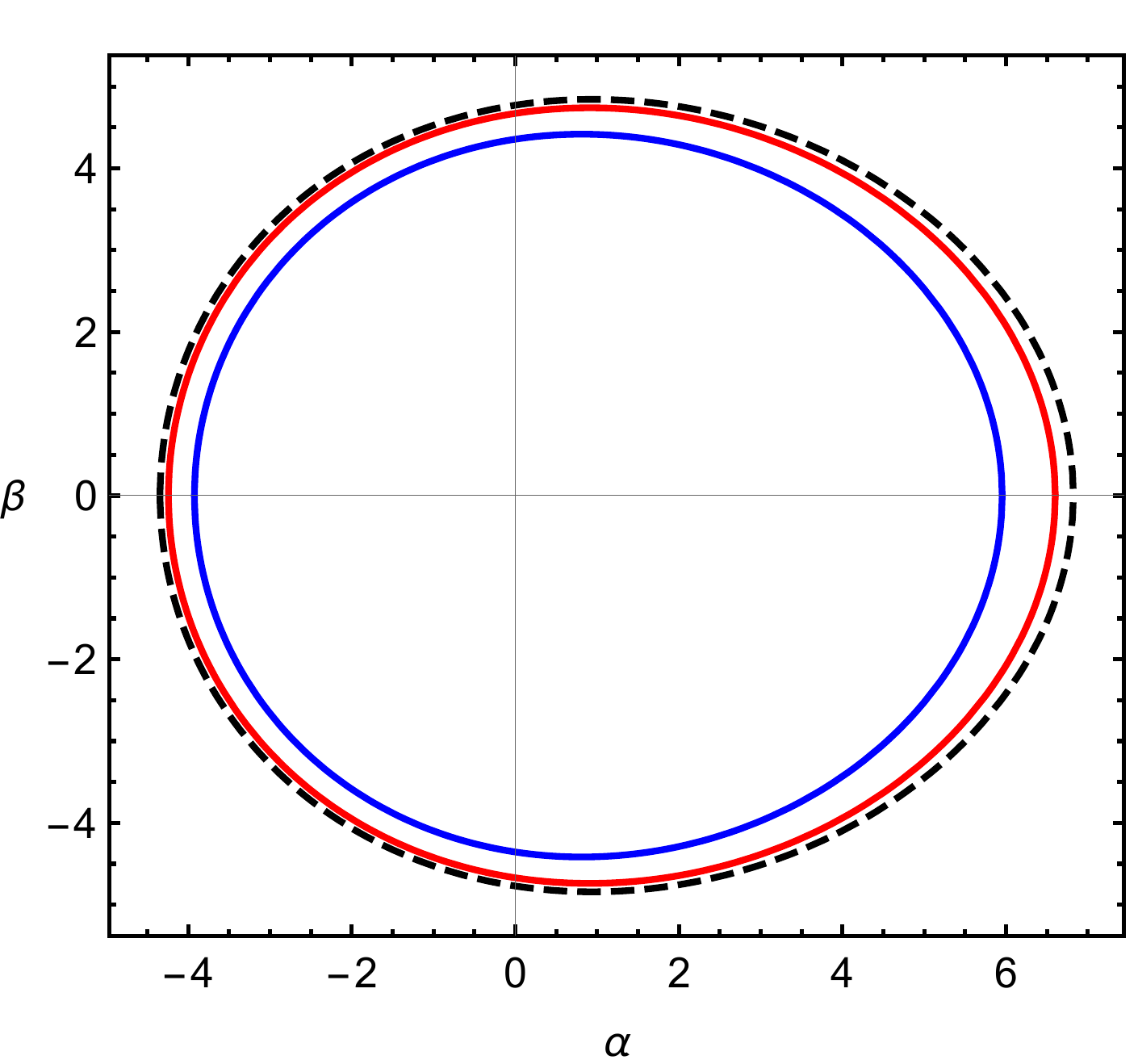} \quad
\includegraphics[scale=0.46]{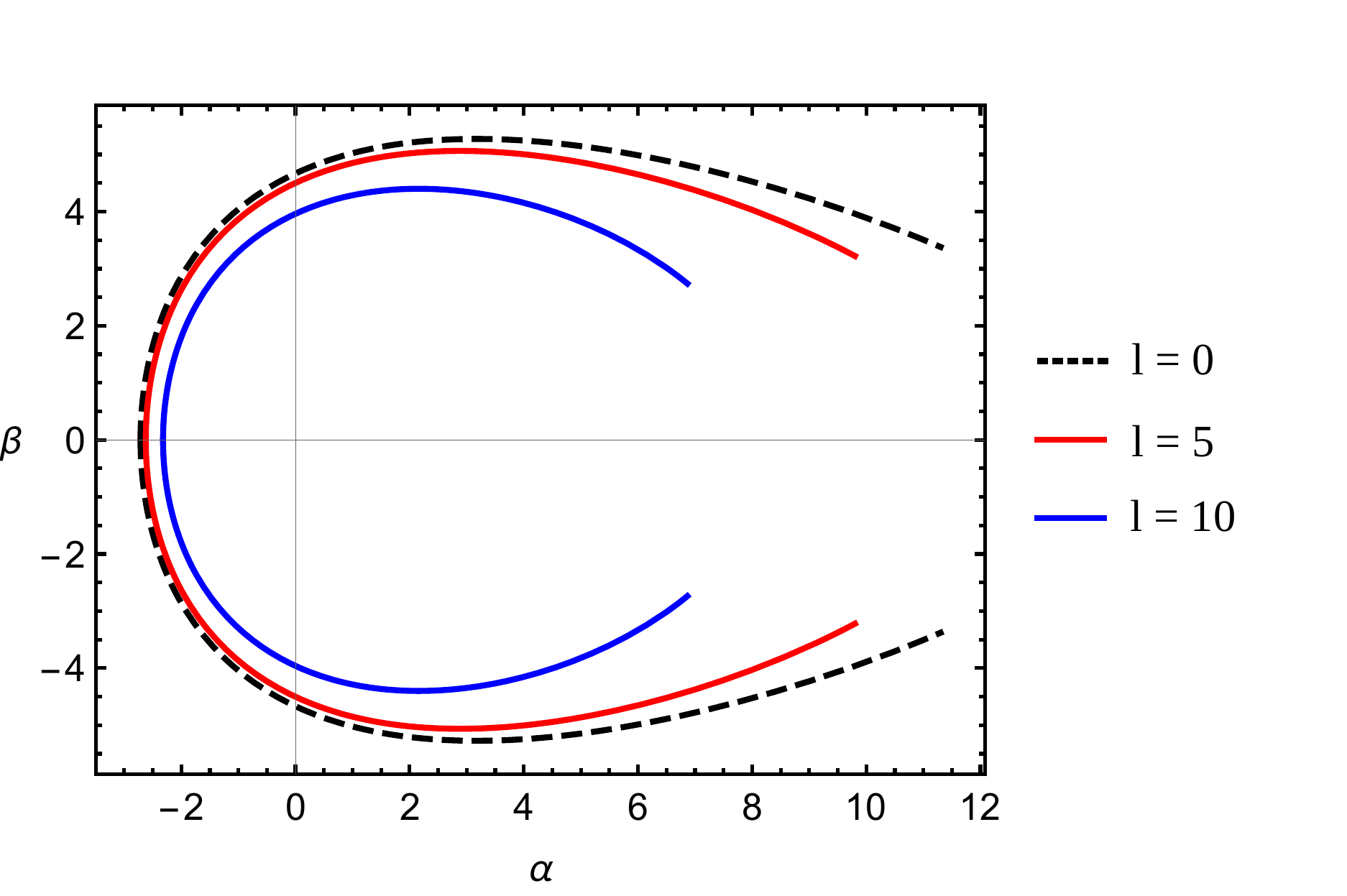} \\
\caption{Critical curves with different $l$. We have set $M=1$,  $a=1$, $b=-0.5$; and the azimuthal angle $\t_o = \f{\pi}{10}$(left) and $ \f{\pi}{2}$(right), the distance $r_o$ is set as $10M$ to show the influence of $l$.}\label{C4}
\end{figure}

Now let us discuss the influence of the parameter $l$. Since the parameter $l$ always appears in the form of $l^2/r_o^2$, a small $l$ has  little effect on the shadow curve. In order to be able to see the change clearly, we vary $l/r_o$ from $0$ to $1$ to show the influence of $l$ on the shadow curve in Fig. \ref{C4}. The plots show that when $l$ gets bigger, the size of the shadow would be smaller.

\begin{figure}[H]
\centering
\includegraphics[scale=0.45]{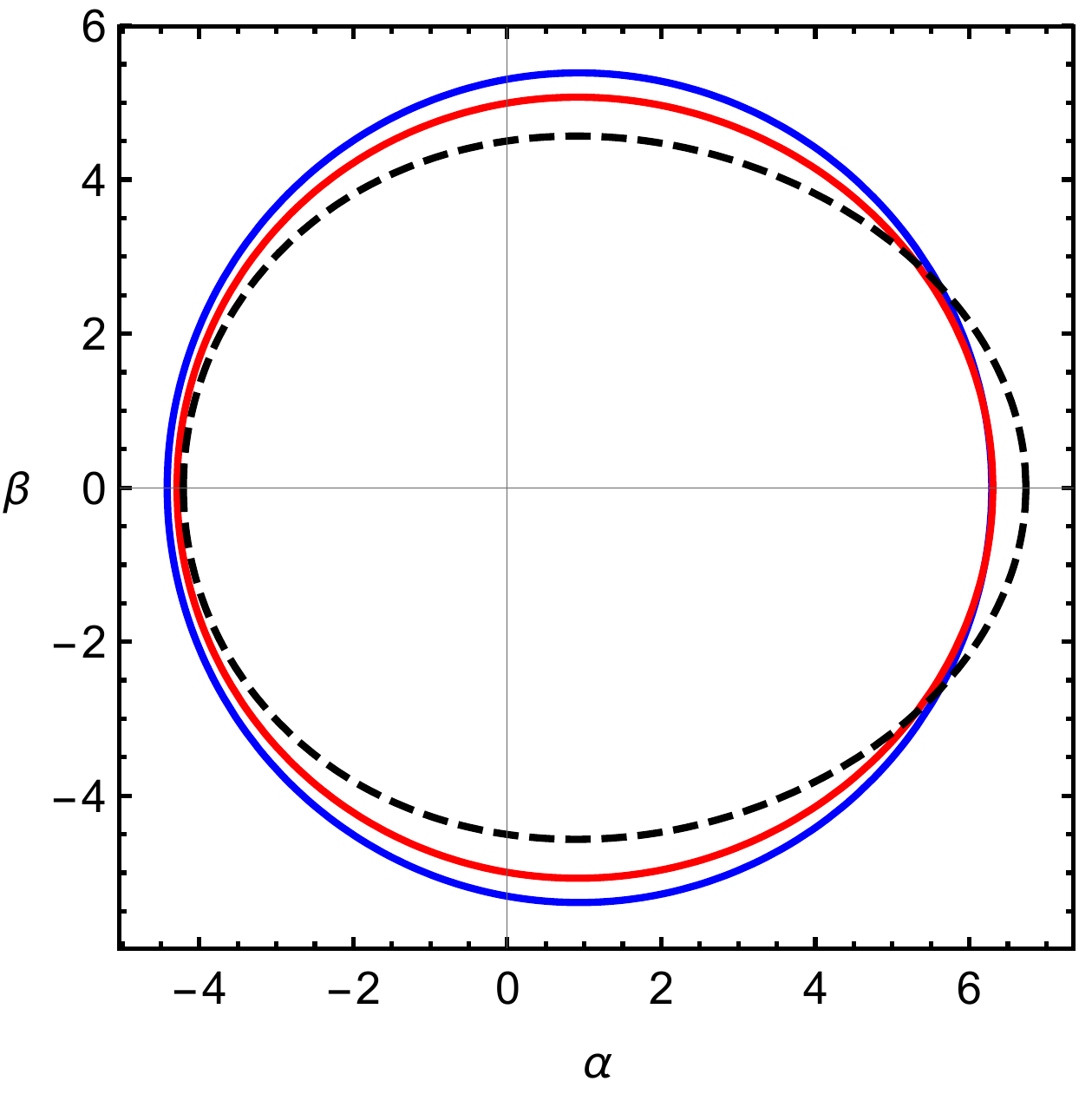} \quad 
\includegraphics[scale=0.45]{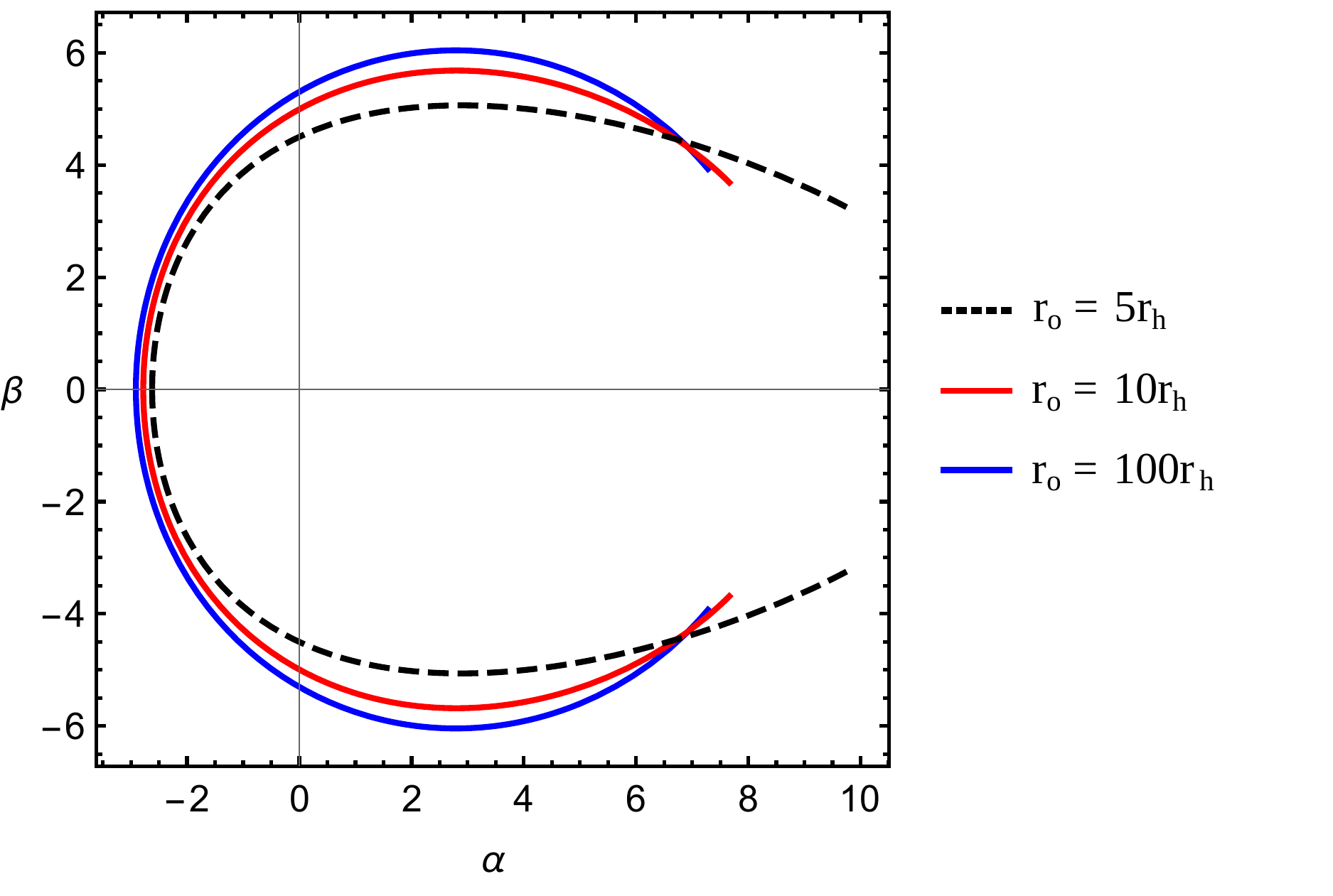} \\
\caption{Critical curves with different $r_o$. We have set $M=1$ ,  $a=1$ , $b=-0.5$, and $l=5$; and the azimuthal angle $\t_o = \f{\pi}{10}$(left) and $ \f{\pi}{2}$(right).}\label{C5}
\end{figure}

At last, we show the variations of the shadow curves with respect to the distance of observers from the black hole. From Fig. \ref{C5}, we find that the shape of the shadow curve is getting rounder with increasing $r_o$. This is because that the drag effect of the spin becomes less pronounced as the distance increases.

\section{Applying to the shadow of M87*}\label{sec4} 

In this section, we move to apply our above results to the observed black hole shadow of M87*. First of all, the shadow curve of M87* is closed which implies the observational angle satisfies
\be
\t_c(a,b) \geq 17^{\circ} \ ,
\ee
or equivalently one finds
\be
 \lm(\td{r}_+) \geq -0.413a \ .
\ee
This condition potentially could give a constraint on the spin and the tidal charge. However, as we have shown, 
within our paramter ranges $0.1<a<1.5$ and $-2<b\le1-a^2$, the maximal and minimal values of $\t_c$ are $0.44\pi$ and $0.32\pi$ respectively, both of which are obviously bigger than $17^{\circ}\sim 0.094\pi$. Thus the critical curve is always closed in this range. In addition, the distance of the observer from the center of M87* is estimated to be $D=(16.8\pm0.8)$ Mpc \cite{Blakeslee:2009tc, Bird:2010rd}, and the mass is considered to be $M=(6.5\pm0.7)\times10^9 M_\odot$ \cite{Akiyama:2019cqa, Akiyama:2019fyp, Akiyama:2019eap}. Note that our results are expressed in Planck units, that is one sets $c=G=1$. Then in our convention, we have $D\sim5\times10^{10}$ which is a huge quantity. Recall that $l$ appears in the form of $l^2/r_o^2$, thus the effects of $l$ can be neglected when considering the shadow curve of M87*. Therefore, in the following we only focus on the effects of the spin $a$ and the tidal charge $b$.

\begin{figure}[H]
\centering
\includegraphics[scale=0.3]{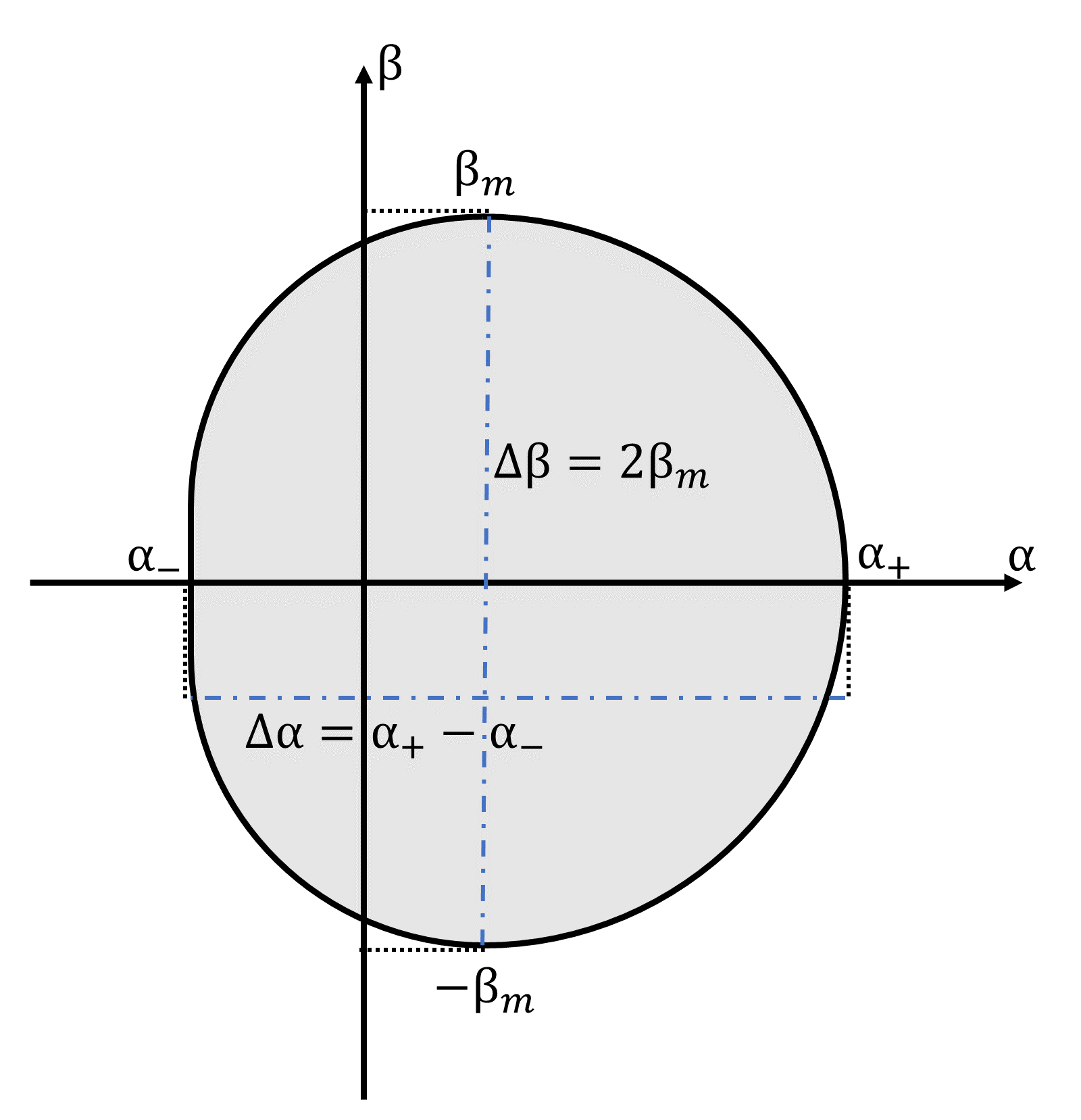}\\
\caption{ }\label{o1}
\end{figure}

Since the shadow is not a perfect circular in general, as seen in Fig. \ref{o1}, the major axis $\D\b$ and the minor axis $\D\a$ should be introduced (as can be seen in \cite{Kumar:2018ple}) and the axis ratio can be defined as 
\be
\delta=\frac{\Delta\beta}{\Delta\alpha}
\ee
which is supposed to be smaller than $4/3$ for M87* black hole shadow \cite{Akiyama:2019cqa, Akiyama:2019fyp, Akiyama:2019eap}.

\begin{figure}[H]
\centering
\includegraphics[scale=0.32]{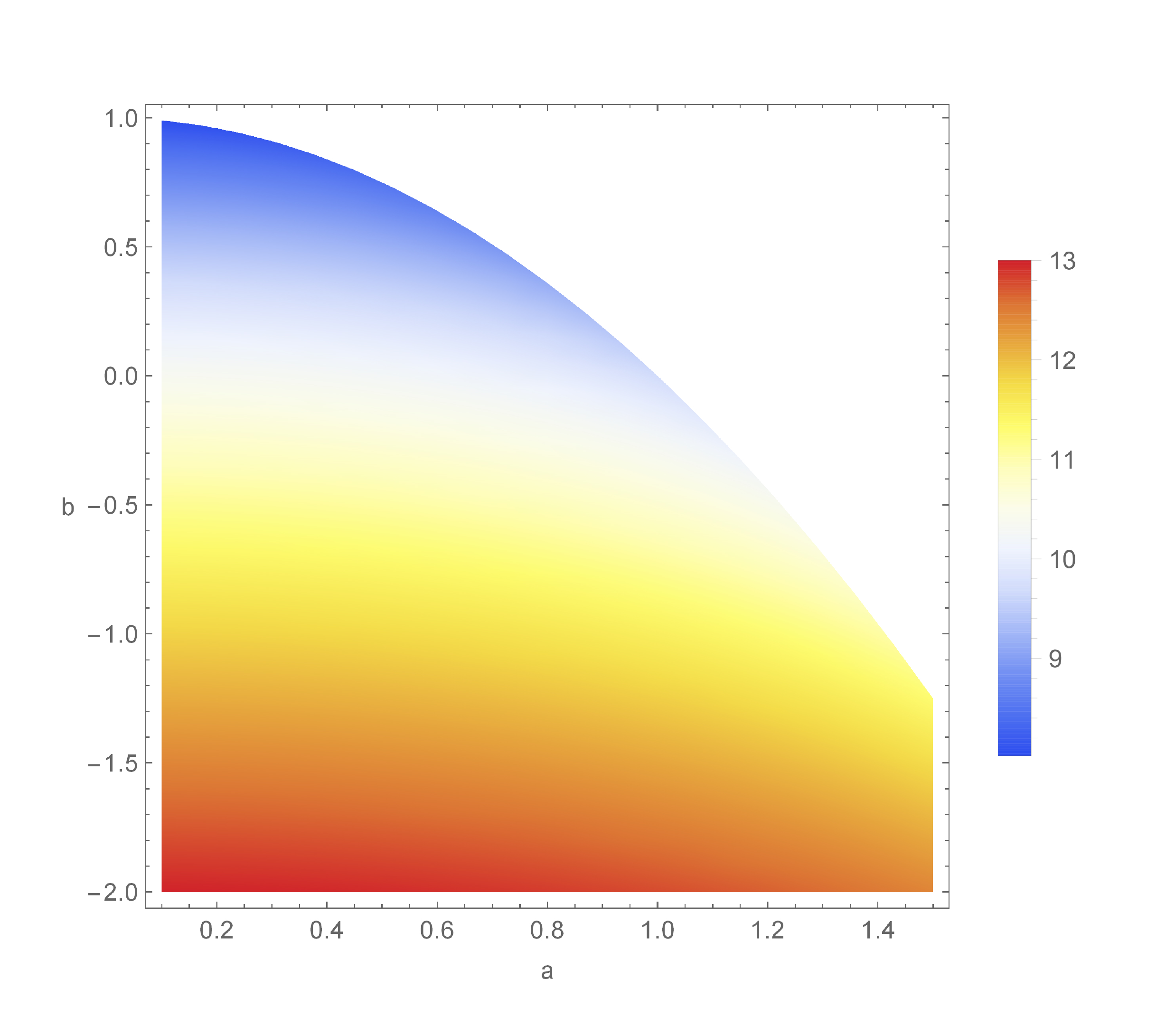}\quad\includegraphics[scale=0.32]{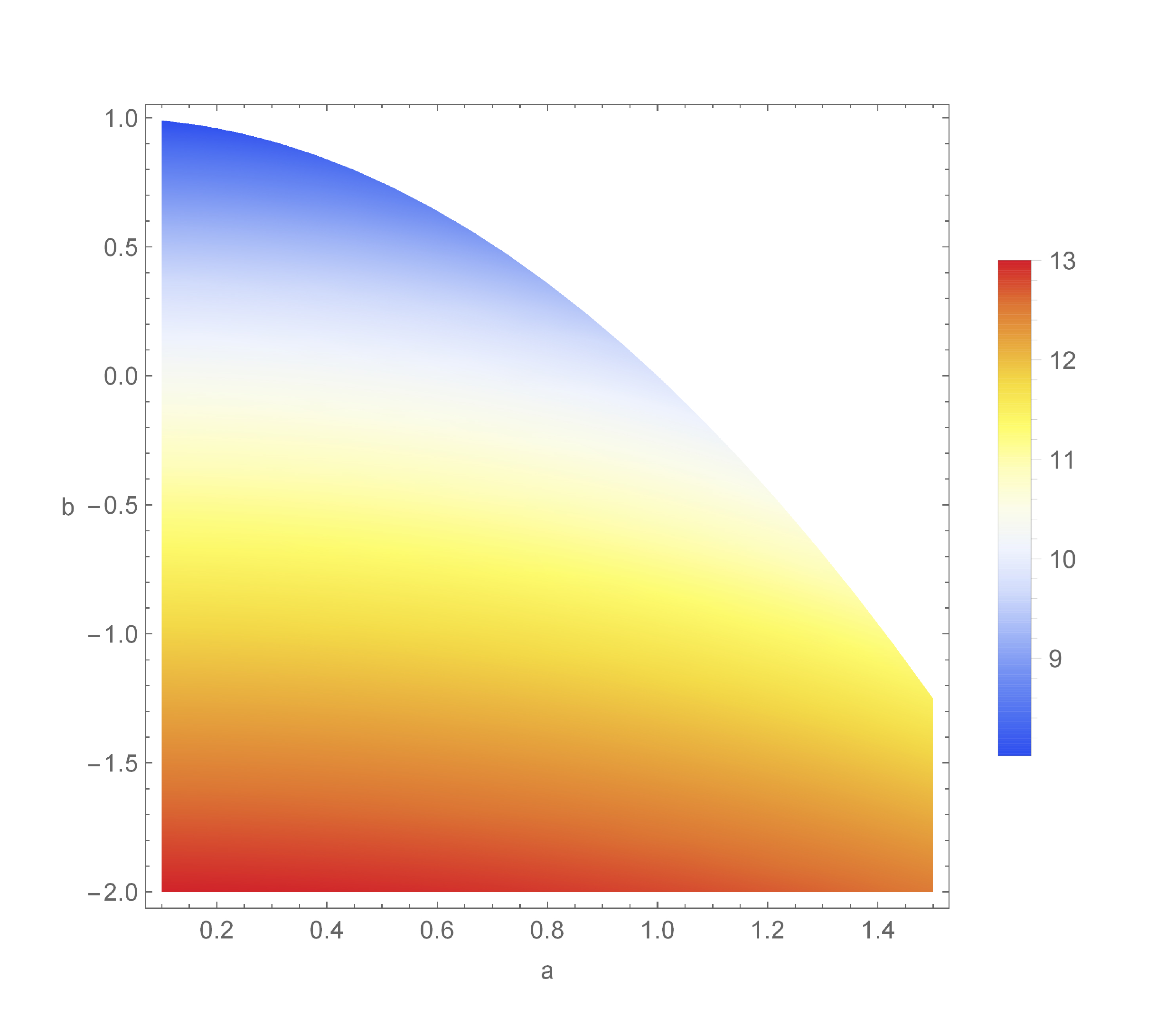} \\
\caption{Density plot of $\D\a$ and $\D\b$. The left figure is for $\D \a$, while the right for $\D \b$. }\label{D}
\end{figure}

In Fig. \ref{D}, we show the density plot of $\Delta \alpha$ and $\Delta \beta$ with respect to the spin $a$ and tidal charge $b$. We can see that the length of $\Delta\alpha$ and $\Delta\beta$ increase with an increasing $a$ or decreasing $b$.

\begin{figure}[H]
\centering
\includegraphics[scale=0.5]{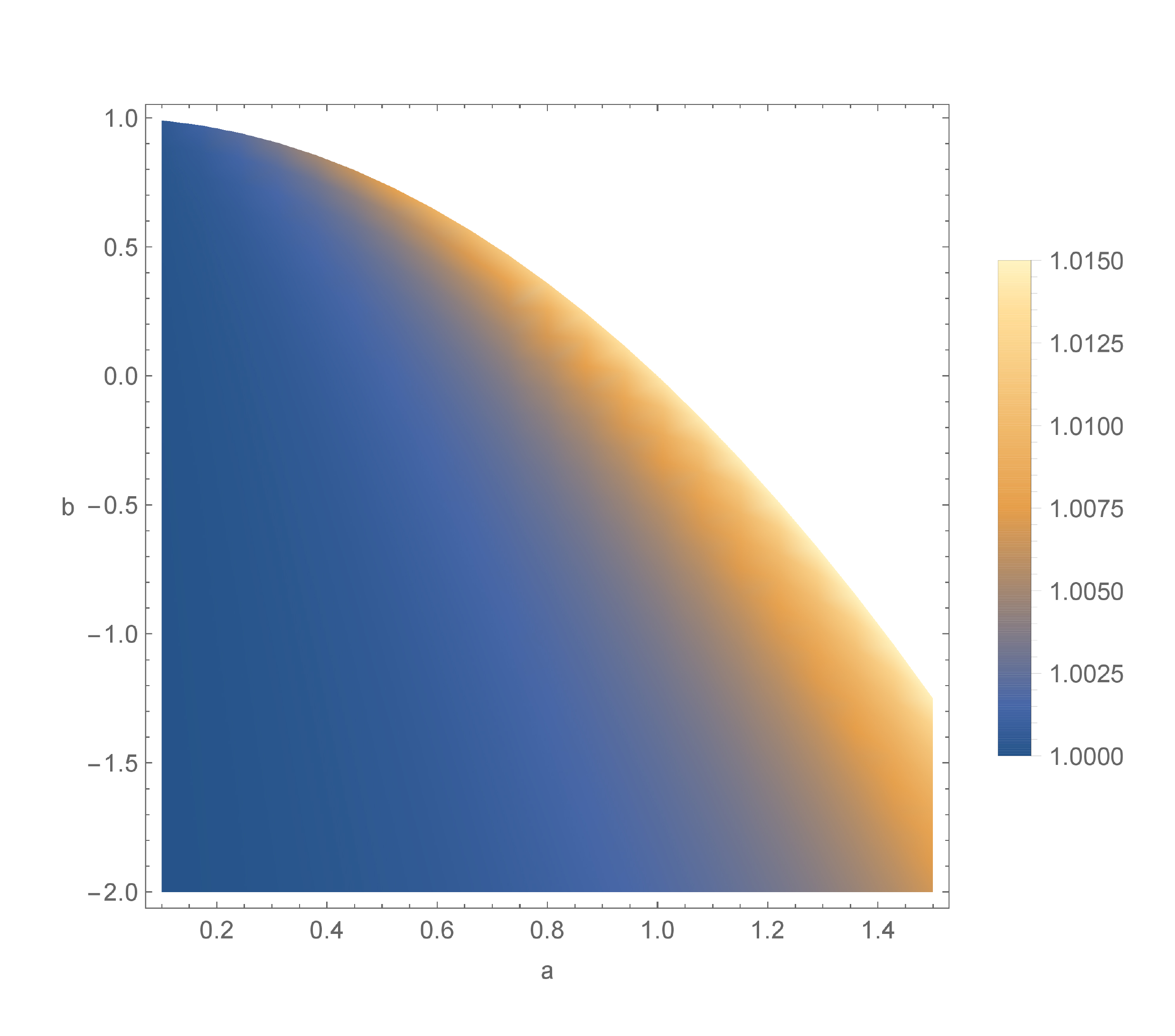}\\
\caption{Density plot of $\delta=\D\b/\D\a$.}\label{DbDa}
\end{figure}

In Fig. \ref{DbDa}, we present the density plot of the axis ratio $\delta$ with respect to $a$ and $b$. Obviously, from the graph we can see $\delta>1$ always holds and the maximum $\delta_{max}=1.015<4/3$, that is, the observational data of M87* black hole shadow cannot rule out the existence of braneworld black hole spacetime in this sense. 

\begin{figure}[H]
\centering
\includegraphics[scale=0.5]{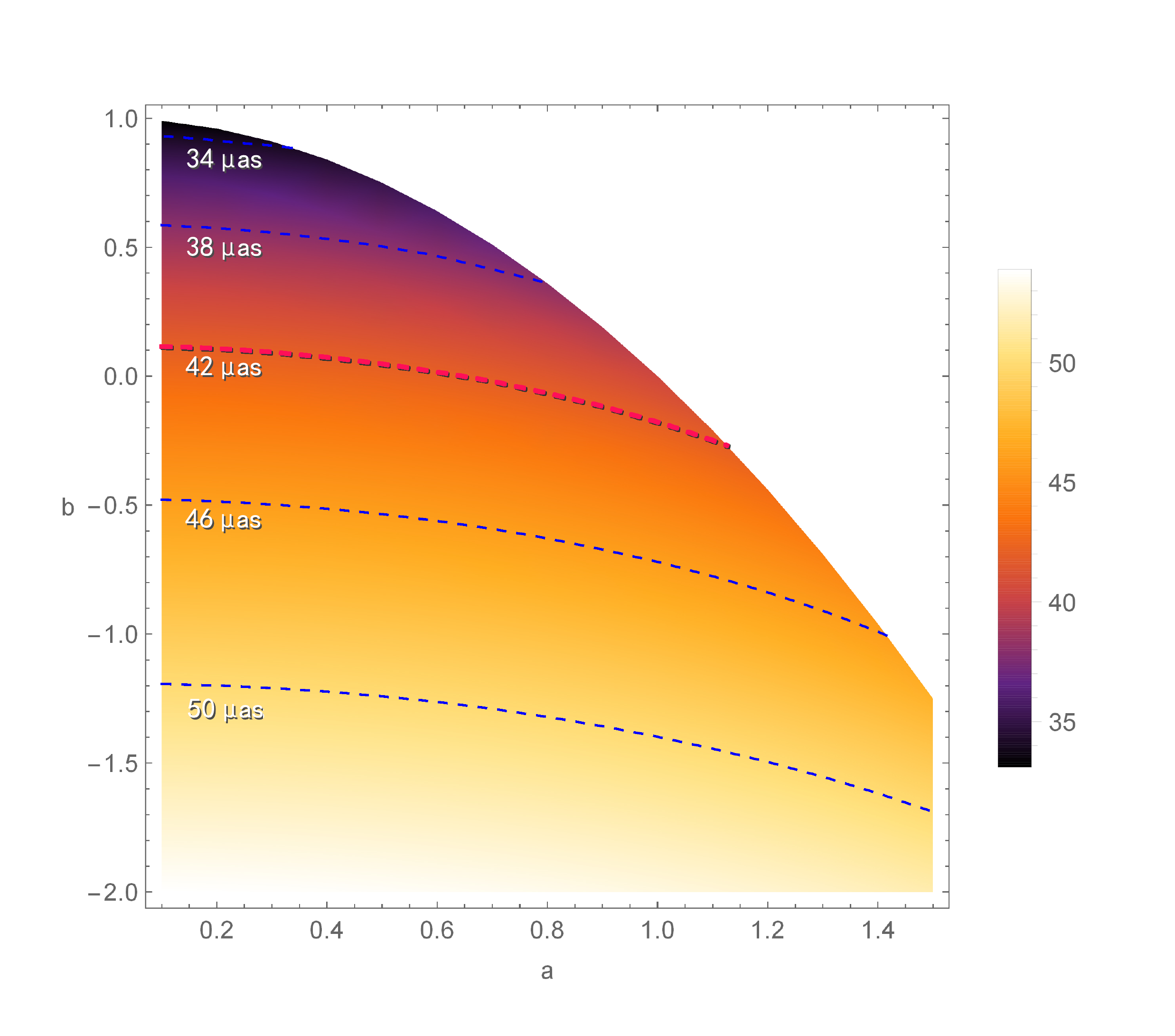}\\
\caption{Density plot of $\theta_{sh}=R_{av}/D$, here we use $M=6.5\times10^9M_{\odot}$.}\label{AngularDiameter1}
\end{figure}

Another important quantity representing the geometric shape is the average radius $R_{av}$ which is defined by 
\bea
R_{av} \equiv \sqrt{\f{Area}{\pi}}.
\eea
Here $Area =2\int_{\a_-}^{\a_+}\beta d\a$ is the area of shadow. We wanted to emphasize that in \cite{BSS} the authors define the average radius using polar coordinates. The angular diameter is defined as $\theta_{sh}=R_{av}/D$, then we give the density plot of the angular diameter with respect to $a$ and $b$ in Fig. \ref{AngularDiameter1}. Also, we draw $34\mu as$,  $38\mu as$,  $42\mu as$,  $46\mu as$ and $50\mu as$ lines in the density plot.

As we know, the best estimate of angular diameter is $42\mu as$ and the spin is estimated to be $0.5\le|a|\le0.94$ from the EHT collaboration. From Fig. \ref{AngularDiameter1}, along the $42\mu as$ line we find that $b\sim0.05$ for $a=0.5$ and $b\sim-0.14$ for $a=0.94$. Compared to the results in \cite{BSS}, which find $b=-0.4$ for $M=6.5\times10^9M_{\odot}$, our results give a larger $b$ for the same $a$, and the difference is distinct and observable. In addition, the analysis in \cite{BSS} favors a negative $b$ while our results supports a positive $b$ when $a=0.5$. Besides, for a high spin, even though our results also favor a negative $b$, the strength of the tidal charge we found is much smaller, which means that our treatment shows the deviation from the Kerr black hole is smaller than that in \cite{BSS}.

\section{Closing remarks}\label{sec5}
In this paper, we revisited the shadows of braneworld black holes in the Randall-Sundrum model. The metrics both in the near and the far regions were taken into account. Through a careful analysis of the angular potential functions in both regions, we found the critical curve of impact parameters $\tilde{\eta}(\tilde{\lambda})$ and the lower bound $\eta(\lambda)$ can have one or two intersection points. This fact is different from that in Kerr black hole spacetime. It is also different from the existing studies on the braneworld black hole spacetime using only the near horizon metric. Moreover, as the lower bound $\eta(\lambda)$ is connected with the parameter of the shadow curve $\beta$, one intersection between $\tilde{\eta}(\tilde{\lambda})$ and $\eta({\lambda})$ means the shadow curve is not closed, while  two intersections means a closed one. Furthermore, we found the critical observational angle $\theta_c$ suggesting that the shadow curve is closed when $0<\theta_o\le\theta_c$ and the shadow curve would be open when $\theta_c<\theta_o\le\pi/2$.

We also studied the influences of $a$, $b$, $l$ and $r_o$ on the shadow curve, and the conclusions can be found in sec. \ref{shadow}. Then we applied our results to the shadow of M87* and calculated diameters and angular diameters of the shadow curve. Our density plots showed that  our results favor a larger $b$ for the fixed $a$, compared with the ones in \cite{BSS}.

\section*{Acknowledgments} 
We are very grateful to Bin Wang for the question, which lead to this project, and valuable suggestions and comments on the  manuscript. We would like to thank Peng-cheng Li for his participation in the early stage of the project. 
The work is in part supported by NSFC Grant  No. 11735001. MG is also funded by China Postdoctoral Science Foundation Grant No. 2019M660278 and 2020T130020.


\begin{thebibliography}{10}
 
\bibitem{Abbott:2016blz}
B.~P.~Abbott \textit{et al.} [LIGO Scientific and Virgo],
Phys. Rev. Lett. \textbf{116} (2016) no.6, 061102
doi:10.1103/PhysRevLett.116.061102
[arXiv:1602.03837 [gr-qc]].

\bibitem{Abbott:2016nmj}
B.~P.~Abbott \textit{et al.} [LIGO Scientific and Virgo],
Phys. Rev. Lett. \textbf{116} (2016) no.24, 241103
doi:10.1103/PhysRevLett.116.241103
[arXiv:1606.04855 [gr-qc]].

\bibitem{TheLIGOScientific:2017qsa}
B.~P.~Abbott \textit{et al.} [LIGO Scientific and Virgo],
Phys. Rev. Lett. \textbf{119} (2017) no.16, 161101
doi:10.1103/PhysRevLett.119.161101
[arXiv:1710.05832 [gr-qc]].
 
\bibitem{Akiyama:2019cqa}
K.~Akiyama \textit{et al.} [Event Horizon Telescope],
Astrophys. J. \textbf{875} (2019) no.1, L1
doi:10.3847/2041-8213/ab0ec7
[arXiv:1906.11238 [astro-ph.GA]].
 
\bibitem{Akiyama:2019brx}
K.~Akiyama \textit{et al.} [Event Horizon Telescope],
Astrophys. J. Lett. \textbf{875} (2019) no.1, L2
doi:10.3847/2041-8213/ab0c96
[arXiv:1906.11239 [astro-ph.IM]].
 
\bibitem{Akiyama:2019sww}
K.~Akiyama \textit{et al.} [Event Horizon Telescope],
Astrophys. J. Lett. \textbf{875} (2019) no.1, L3
doi:10.3847/2041-8213/ab0c57
[arXiv:1906.11240 [astro-ph.GA]].
 
\bibitem{Akiyama:2019bqs}
K.~Akiyama \textit{et al.} [Event Horizon Telescope],
Astrophys. J. Lett. \textbf{875} (2019) no.1, L4
doi:10.3847/2041-8213/ab0e85
[arXiv:1906.11241 [astro-ph.GA]].
 
\bibitem{Akiyama:2019fyp}
K.~Akiyama \textit{et al.} [Event Horizon Telescope],
Astrophys. J. Lett. \textbf{875} (2019) no.1, L5
doi:10.3847/2041-8213/ab0f43
[arXiv:1906.11242 [astro-ph.GA]].
 
\bibitem{Akiyama:2019eap}
K.~Akiyama \textit{et al.} [Event Horizon Telescope],
Astrophys. J. Lett. \textbf{875} (2019) no.1, L6
doi:10.3847/2041-8213/ab1141
[arXiv:1906.11243 [astro-ph.GA]].
 
 \bibitem{BSS}
 I.~Banerjee, S.~Chakraborty, S.~Sengupta,
 Phys.\ Rev.\ D.101.041301 (2020)
  doi:10.1103/PhysRevD.101.041301
 [arXiv: 1909.09385[gr-qc]].
 
\bibitem{Christodoulou:1991yfa}
D.~Christodoulou, 
Commun. Pure Appl. Math. \textbf{44} (1991) no.3, 339-373 doi:10.1002/cpa.3160440305
 
 \bibitem{Penrose:1964wq}
R.~Penrose,
Phys. Rev. Lett. \textbf{14} (1965), 57-59
doi:10.1103/PhysRevLett.14.57

\bibitem{Hawking:1969sw}
S.~W.~Hawking and R.~Penrose,
Proc. Roy. Soc. Lond. A \textbf{314} (1970), 529-548
doi:10.1098/rspa.1970.0021
 
 \bibitem{Milgrom:2003ui}
M.~Milgrom and R.~H.~Sanders,
Astrophys. J. Lett. \textbf{599} (2003), L25-L28
doi:10.1086/381138
[arXiv:astro-ph/0309617 [astro-ph]].
 
\bibitem{Peebles:2002gy}
P.~J.~E.~Peebles and B.~Ratra,
Rev. Mod. Phys. \textbf{75} (2003), 559-606
doi:10.1103/RevModPhys.75.559
[arXiv:astro-ph/0207347 [astro-ph]].

\bibitem{Perlmutter:1998np}
S.~Perlmutter \textit{et al.} [Supernova Cosmology Project],
Astrophys. J. \textbf{517} (1999), 565-586
doi:10.1086/307221
[arXiv:astro-ph/9812133 [astro-ph]].

 \bibitem{Nojiri:2017ncd}
S.~Nojiri, S.~D.~Odintsov and V.~K.~Oikonomou,
Phys. Rept. \textbf{692} (2017), 1-104
doi:10.1016/j.physrep.2017.06.001
[arXiv:1705.11098 [gr-qc]].

\bibitem{Maartens:2003tw}
R.~Maartens,
Living Rev. Rel. \textbf{7} (2004), 7
doi:10.12942/lrr-2004-7
[arXiv:gr-qc/0312059 [gr-qc]].
  
 \bibitem{Kanti:2004nr}
P.~Kanti,
Int. J. Mod. Phys. A \textbf{19} (2004), 4899-4951
doi:10.1142/S0217751X04018324
[arXiv:hep-ph/0402168 [hep-ph]].
  
 \bibitem{Chakravarti:2019aup}
K.~Chakravarti, S.~Chakraborty, K.~S.~Phukon, S.~Bose and S.~SenGupta,
Class. Quant. Grav. \textbf{37} (2020) no.10, 105004
doi:10.1088/1361-6382/ab8355
[arXiv:1903.10159 [gr-qc]]. 

\bibitem{Guo:2020zmf}
M.~Guo and P.~C.~Li,
Eur. Phys. J. C \textbf{80}, no.6, 588 (2020)
doi:10.1140/epjc/s10052-020-8164-7
[arXiv:2003.02523 [gr-qc]].


\bibitem{Konoplya:2020bxa}
R.~A.~Konoplya and A.~F.~Zinhailo,
Eur. Phys. J. C \textbf{80}, no.11, 1049 (2020)
doi:10.1140/epjc/s10052-020-08639-8
[arXiv:2003.01188 [gr-qc]].

\bibitem{Kumar:2020owy}
R.~Kumar and S.~G.~Ghosh,
JCAP \textbf{07}, no.07, 053 (2020)
doi:10.1088/1475-7516/2020/07/053
[arXiv:2003.08927 [gr-qc]].

\bibitem{Wei:2020ght}
S.~W.~Wei and Y.~X.~Liu,
[arXiv:2003.07769 [gr-qc]].

\bibitem{Bisnovatyi-Kogan:2019wdd}
G.~S.~Bisnovatyi-Kogan, O.~Y.~Tsupko and V.~Perlick,
PoS \textbf{MULTIF2019}, 009 (2019)
doi:10.22323/1.362.0009
[arXiv:1910.10514 [gr-qc]].


\bibitem{Bisnovatyi-Kogan:2018vxl}
G.~S.~Bisnovatyi-Kogan and O.~Y.~Tsupko,
Phys. Rev. D \textbf{98}, no.8, 084020 (2018)
doi:10.1103/PhysRevD.98.084020
[arXiv:1805.03311 [gr-qc]].

\bibitem{Li:2020drn}
P.~C.~Li, M.~Guo and B.~Chen,
Phys. Rev. D \textbf{101}, no.8, 084041 (2020)
doi:10.1103/PhysRevD.101.084041
[arXiv:2001.04231 [gr-qc]].
 

\bibitem{Cunha:2017eoe}
P.~V.~P.~Cunha, C.~A.~R.~Herdeiro and E.~Radu,
Phys. Rev. D \textbf{96}, no.2, 024039 (2017)
doi:10.1103/PhysRevD.96.024039
[arXiv:1705.05461 [gr-qc]].
 
\bibitem{Qian:2021qow}
W.~L.~Qian, S.~Chen, C.~G.~Shao, B.~Wang and R.~H.~Yue,
[arXiv:2102.03820 [gr-qc]].

\bibitem{Chen:2020qyp}
S.~Chen, M.~Wang and J.~Jing,
JHEP \textbf{07}, 054 (2020)
doi:10.1007/JHEP07(2020)054
[arXiv:2004.08857 [gr-qc]].

\bibitem{Hu:2020usx}
Z.~Hu, Z.~Zhong, P.~C.~Li, M.~Guo and B.~Chen,
Phys. Rev. D \textbf{103}, no.4, 044057 (2021)
doi:10.1103/PhysRevD.103.044057
[arXiv:2012.07022 [gr-qc]].
  

\bibitem{Wang:2020emr}
X.~Wang, P.~C.~Li, C.~Y.~Zhang and M.~Guo,
Phys. Lett. B \textbf{811}, 135930 (2020)
doi:10.1016/j.physletb.2020.135930
[arXiv:2007.03327 [gr-qc]].


\bibitem{Wielgus:2020uqz}
M.~Wielgus, J.~Horak, F.~Vincent and M.~Abramowicz,
Phys. Rev. D \textbf{102}, no.8, 084044 (2020)
doi:10.1103/PhysRevD.102.084044
[arXiv:2008.10130 [gr-qc]].

\bibitem{Tsukamoto:2021fpp}
N.~Tsukamoto,
[arXiv:2101.07060 [gr-qc]].


\bibitem{Guerrero:2021pxt}
M.~Guerrero, G.~J.~Olmo and D.~Rubiera-Garcia,
[arXiv:2102.00840 [gr-qc]].


\bibitem{Peng:2021osd}
J.~Peng, M.~Guo and X.~H.~Feng,
[arXiv:2102.05488 [gr-qc]].
  
\bibitem{Amarilla:2011fx}
L.~Amarilla and E.~F.~Eiroa,
Phys. Rev. D \textbf{85}, 064019 (2012)

\bibitem{Eiroa:2017uuq}
E.~F.~Eiroa and C.~M.~Sendra,
Eur. Phys. J. C \textbf{78}, no.2, 91 (2018)

\bibitem{Abdujabbarov:2017pfw}
A.~Abdujabbarov, B.~Ahmedov, N.~Dadhich and F.~Atamurotov,
Phys. Rev. D \textbf{96}, no.8, 084017 (2017)
  
\bibitem{RWW}
Richard Whisker,
 [arXiv: 0810.1534[gr-qc]].

\bibitem{Aliev:2005bi}
A.~N.~Aliev and A.~E.~Gumrukcuoglu,
Phys. Rev. D \textbf{71} (2005), 104027
doi:10.1103/PhysRevD.71.104027
[arXiv:hep-th/0502223 [hep-th]].
  
\bibitem{Aliev:2009cg}
A.~N.~Aliev and P.~Talazan,
Phys. Rev. D \textbf{80} (2009), 044023
doi:10.1103/PhysRevD.80.044023
[arXiv:0906.1465 [gr-qc]].
  
\bibitem{Dadhich:2000am}
N.~Dadhich, R.~Maartens, P.~Papadopoulos and V.~Rezania,
Phys. Lett. B \textbf{487} (2000), 1-6
doi:10.1016/S0370-2693(00)00798-X
[arXiv:hep-th/0003061 [hep-th]].
  
\bibitem{Gralla:2019ceu}
S.~E.~Gralla and A.~Lupsasca,
Phys. Rev. D \textbf{101} (2020) no.4, 044032
doi:10.1103/PhysRevD.101.044032
[arXiv:1910.12881 [gr-qc]].
  
\bibitem{Bardeen:1973tla}
J.~M.~Bardeen,
C.~Witt and B.~Witt, \ editors, \ Black Holes, \ pp.215, 1973. 
 
\bibitem{Gralla:2017ufe}
S.~E.~Gralla, A.~Lupsasca and A.~Strominger,
Mon. Not. Roy. Astron. Soc. \textbf{475}, no.3, 3829-3853 (2018)

 
\bibitem{Guo:2018kis}
M.~Guo, N.~A.~Obers and H.~Yan,
Phys. Rev. D \textbf{98} (2018) no.8, 084063
doi:10.1103/PhysRevD.98.084063
[arXiv:1806.05249 [gr-qc]].
 

\bibitem{Yan:2019etp}
H.~Yan,
Phys. Rev. D \textbf{99}, no.8, 084050 (2019)


 
\bibitem{Guo:2019lur}
M.~Guo, S.~Song and H.~Yan,
Phys. Rev. D \textbf{101} (2020) no.2, 024055
doi:10.1103/PhysRevD.101.024055
[arXiv:1911.04796 [gr-qc]].


\bibitem{Porfyriadis:2016gwb}
A.~P.~Porfyriadis, Y.~Shi and A.~Strominger,
Phys. Rev. D \textbf{95}, no.6, 064009 (2017)


 
\bibitem{Li:2020val}
P.~C.~Li, M.~Guo and B.~Chen,
Class. Quant. Grav. \textbf{38} (2021) no.6, 065008
doi:10.1088/1361-6382/abd860
[arXiv:2006.05153 [gr-qc]].
 

\bibitem{Guo:2019pte}
M.~Guo, P.~C.~Li and B.~Chen,
Phys. Rev. D \textbf{101}, no.2, 024054 (2020)

 
 
\bibitem{Blakeslee:2009tc}
J.~P.~Blakeslee, A.~Jordan, S.~Mei, P.~Cote, L.~Ferrarese, L.~Infante, E.~W.~Peng, J.~L.~Tonry and M.~J.~West,
Astrophys. J. \textbf{694} (2009), 556-572
doi:10.1088/0004-637X/694/1/556
[arXiv:0901.1138 [astro-ph.CO]].
  
\bibitem{Bird:2010rd}
S.~Bird, W.~E.~Harris, J.~P.~Blakeslee and C.~Flynn,
Astron. Astrophys. \textbf{524} (2010), A71
doi:10.1051/0004-6361/201014876
[arXiv:1009.3202 [astro-ph.GA]].
  
\bibitem{Kumar:2018ple}
R.~Kumar and S.~G.~Ghosh,
Astrophys. J. \textbf{892} (2020), 78
doi:10.3847/1538-4357/ab77b0
[arXiv:1811.01260 [gr-qc]].

  
  
\end{thebibliography}
\end{document}